\documentclass[pdflatex,sn-mathphys-num]{sn-jnl}

\usepackage{graphicx}%
\usepackage{multirow}%
\usepackage{amsmath,amssymb,amsfonts}%
\usepackage{amsthm}%
\usepackage{mathrsfs}%
\usepackage[title]{appendix}%
\usepackage{xcolor}%
\usepackage{textcomp}%
\usepackage{manyfoot}%
\usepackage{booktabs}%
\usepackage{algorithm}%
\usepackage{algorithmicx}%
\usepackage{algpseudocode}%
\usepackage{listings, comment}%
\graphicspath{{figures/}}

\usepackage{caption}
\usepackage{subcaption}
\usepackage{physics}
\usepackage{gensymb}
\usepackage{xurl}

\newgeometry{left=2cm,right=2cm, top=2cm, bottom=2cm}   

\AtBeginDocument{}

\theoremstyle{thmstyleone}%
\theoremstyle{thmstyletwo}%

\theoremstyle{thmstylethree}%

\begin{document}

\title[Article Title]{Transferable Implicit Solvent Machine Learning Potential for Drugs and Proteins Approaching Ab Initio Accuracy}

\author[1]{\fnm{Jan} \sur{Eckwert}}
\author*[1,2]{\fnm{Julija} \sur{Zavadlav}}\email{julija.zavadlav@tum.de}

\affil[1]{\orgdiv{Multiscale Modeling of Fluid Materials, Department of Engineering Physics and
		Computation, TUM School of Engineering and Design and Department of Physics, TUM School of Natural Sciences}, \orgname{Technical University of Munich}, \orgaddress{\city{Munich}, \postcode{80333}, \country{Germany}}}

\affil[2]{\orgdiv{Atomistic Modeling Center}, \orgname{Munich Data Science Institute, Technical University of Munich}, \orgaddress{\city{Garching}, \postcode{85748}, \country{Germany}}}

\abstract{
    Machine learning interatomic potentials (MLPs) have revolutionized atomistic modeling, offering the potential to replace traditional methods like Density Functional Theory (DFT). However, inference time of MLPs is orders of magnitude slower than that of classical force fields, hindering real-world applications for biomolecular systems that require timescales of microseconds and beyond. Implicit solvent MLPs can address this issue, but are faced with data challenges associated with coarse-grained modeling. Consequently, previous approaches relied on empirical force field data, thereby inherently limiting the MLP's accuracy. Here, we introduce the Transferable Water Implicit Network (TWIN), an implicit water MLP parametrized entirely by an Equivariant Graph Neural Network and trained solely on ab initio and experimental labels. We demonstrate TWIN's transferability across drug-like molecules, peptides, and proteins, achieving excellent results on ab initio and experimental crystallographic and NMR benchmarks, consistently outperforming previous machine-learning-based implicit solvent or coarse-grained models. Furthermore, TWIN closely matches DFT-based explicit solvent MLPs while providing a two-order-of-magnitude faster timestep evaluation, paving the way for efficient ab initio-level modeling of biomolecular systems in aqueous environments. 
	}


\maketitle

\section*{Introduction}\label{sec:introduction}

Understanding protein dynamics is essential for elucidating the molecular mechanisms underlying biological function. Functional processes, such as transitions between metastable conformational states, allosteric regulation, and protein-drug interactions, are governed by the exploration of complex conformational ensembles \cite{karplus_molecular_2002, BiomolecularSimulation_Dror_2012, MolecularDynamics_Scott_2018}. Accurately predicting these ensembles and their interactions with drug-like molecules is therefore critical for modern medicine and molecular-level understanding of life \cite{durrant_molecular_2011, DeVivo_RoleofMolecularDynamics_2016}. The atomistic description of biologically relevant protein systems, which often encompass tens to hundreds of thousands of atoms, remains computationally prohibitive for ab initio electronic‑structure methods such as density functional theory (DFT). As a result, all atom molecular dynamics (MD) simulations employing classical empirical force fields are the main workhorse for modeling protein conformations, conformational transitions, and protein-ligand interactions \cite{Ponder_ForceFieldsforProteinSimulations_2003, Mackerell_Empiricalforcefieldsforbiological_2004, Lopes_CurrentStatusofProteinForceFields_2025}. 
While these models provide the computational efficiency necessary to treat large biophysical systems, their reliance on empirical parameterizations imposes inherent limitations on their predictive capabilities \cite{Spoel_Systematicdesignofbiomolecularforcefields_2021, kabylda_atomsinteractmolecules_2026}.

The emergence of Machine Learning Potentials (MLPs) has fundamentally transformed atomistic modeling, enabling simulations of million-atom systems with accuracies approaching those of quantum‑mechanical methods~\cite{stoermer_aluminumsolidificationnanopolycrystaldeformation_2026, Fuchs_chemtrainDeploy_2025}. Symmetry‑equivariant message passing graph neural network architectures such as MACE~\cite{Batatia_mace_2022} have been central to this progress, as they provide complex representations of molecular potential‑energy surfaces while rigorously respecting the underlying symmetries of atomic systems. In parallel, the availability of large, diverse quantum‑chemical datasets has substantially expanded the applicability of these models across broad regions of chemical space~\cite{levine_omol_2025, kabylda_qcell_2026}. A prominent example is the SPICE dataset~\cite{eastman_spice_2023}, which provides DFT labels for a broad spectrum of drug‑like molecules, biomolecular fragments, solvated species, and non‑covalent complexes, thereby capturing chemical environments highly relevant to protein and drug modeling. Together, these advances have accelerated the development of transferable, foundation‑like interatomic potentials such as MACE-OFF capable of describing both intra‑ and intermolecular interactions in complex biomolecular systems~\cite{kovacs_2025_maceoff, Kabylda_so3lr_2025, Unke_gems_2024}. Nevertheless, biophysical simulations pose challenges that extend beyond spatial scale alone, requiring access to long (microsecond and beyond) timescales that remain computationally demanding even for state‑of‑the‑art atomistic MLPs. Although enhanced‑sampling techniques can partially mitigate these limitations, their effectiveness often depends on the identification of suitable collective variables, which remains a non-trivial and active area of ongoing research~\cite{henin_enhanced_2022, Mehdi_EnhancedSampling_2024, Zhu_EnhancedSampling_2026}.

To overcome computational limitations of all-atom simulations, the implicit solvent or, more generally, coarse-grained (CG) approaches reduce the number of degrees of freedom, thereby improving efficiency and enabling longer effective sampling~\cite{Noid_gbbio_2013, Singh_cgmodels_2019, Airas_transferableIsGNN, Ding_clcgff_2022}. Classical implicit solvent methods typically rely on generalized Born models, in which the solvent is represented as a dielectric continuum and the solvation free energy is decomposed into electrostatic and nonpolar contributions~\cite{Onufriev_gnnbis_2019}. Although these approaches can provide substantial speedups, they often fail to reproduce the structural and dynamical behavior observed in explicit solvent atomistic simulations~\cite{chen_machine_2021}.
In this context, MLPs offer a flexible route to more accurate reduced representations by learning many-body potentials of mean force (PMF) associated with integrated-out solvent or atomistic degrees of freedom~\cite{majewski_machine_2023, Wang_mlcgff_2019, Noid_RigorousProgressinCoarseGraining_2024,Slejko2026,Coste2023, Thaler_REmin_2022}. 
Such models can capture solvent effects or CG effective interactions while maintaining closer agreement with atomistic reference simulations and generalizing beyond (macro)molecules encountered during training. Recent examples illustrate this potential. The GNNIS model uses a graph neural network-based parameterization of the generalized Born implicit solvent model, which, in combination with a classical force field for the solute, demonstrates transferability across diverse drug-like molecules~\cite{katzberger_2024_ageneralgraph}. Conversely, CGSchNet model~\cite{charron_2023_navigating}, which is built on top of an empirical 
CG prior potential, achieved generalizability across various proteins. Traditional solute modeling and prior potentials should promote faster learning and better generalization, especially in the nonphysical configurational regions. However, they also limit potential applications, as a consistent solute/prior parametrization must be available a priori. 

Most existing implicit solvent or CG MLPs, including GNNIS and CGSchNet, are trained on data generated with classical explicit solvent force fields, which fundamentally limits their theoretically achievable accuracy. This reliance stems from the intrinsic differences between bottom-up atomistic and CG parametrizations. Since the mapping from atomistic to CG degrees of freedom is non-injective, the learning target is not a well-defined potential-energy surface but an effective many-body PMF, inferred from noisy atomistic force observations~\cite{Noid_RigorousProgressinCoarseGraining_2024}. As a consequence, bottom-up training of CG MLPS requires orders of magnitude more training data than atomistic models~\cite{Thaler_REmin_2022, durumeric_learning_2026}. For example, CGSchNet~\cite{charron_2023_navigating} used a substantial training dataset comprising 100~$\mu$s of unbiased all-atom MD simulations of folded protein domains from the CATH database, augmented by over 1~$\mu$s of mono- and dipeptide dimer simulation data. Constructing an equivalent dataset using atomistic MLPs or even first principles electronic structure methods, such as DFT, is computationally infeasible and unlikely to become available in the foreseeable future. While biased datasets or unconventional training routines can significantly reduce data requirements, they raise other issues, such as increased computational cost~\cite{Thaler_REmin_2022}, the need for collective variable selection~\cite{Chen_EnhancedSampling_2025}, careful sampling of the reference system~\cite{ding2022contrastive,airas2023transferable}, and selecting the appropriate amount of added configurational noise~\cite{durumeric_learning_2026}. 

An alternative approach to developing MLP-based implicit solvent models is to generate training data using established quantum mechanics-based continuum-solvent methods, such as the conductor-like screening model for real solvents (COSMO-RS)~\cite{klamt1995conductor} and polarizable continuum model (PCM) \cite{Barone_pcm_2013}, or the Solvation Model based on Density (SMD)~\cite{Hou_smd_2010}. 
Unfortunately, large-scale datasets of this type are not available. Moreover, because these models are themselves parametrized against experimental thermodynamic observables, learning an MLP to reproduce their outputs may introduce an unnecessary intermediate step. A more direct strategy is to parametrize the MLP top-down using the same underlying thermodynamic data, for example, experimental hydration free energies. 
This approach allows the model to learn effective solvent-mediated interactions that are directly constrained by experimentally relevant observables rather than by the approximations of an existing continuum solvent model. Top-down training on molecular solvation data has proven to be an effective and data-efficient method for small organic molecules~\cite{roecken_2024_predictingsolvationFE}. However, this approach alone is not sufficient to achieve transferability to significantly larger and more heterogeneous systems, such as proteins and protein-ligand complexes.

In this work, we present a Transferable Water Implicit Network (TWIN) MLP model and showcase its generalizability across drug-like molecules, peptides, proteins, and protein-drug complexes, substantially broadening the chemical space covered by previous ML-based implicit solvent or CG models. TWIN 
does not rely on topology-dependent prior potentials encoding bonded and nonbonded solute interactions. Instead, the water-mediated effective solute interactions are learned from data using a graph neural network architecture. 
Our multiscale training approach exploits a large-scale DFT dataset derived from SPICE, an all-atom MLP, and experimental hydration free energies from the aqueous subset of the CombiSolv dataset, thereby avoiding direct reliance on empirical force-field labels. 
As a result, TWIN outperforms previous ML-based models on standard benchmarks against ab initio and experimental crystallographic and nuclear magnetic resonance (NMR) spectroscopy data. Furthermore, it achieves excellent quantitative agreement with its all-atom MLP counterpart, demonstrating near ab initio accuracy.

\section*{Results}\label{sec:results}

\subsection*{Transferable Water Implicit Network (TWIN) model}

\begin{figure*}[h!]
	\centering
	\includegraphics[width=\textwidth]{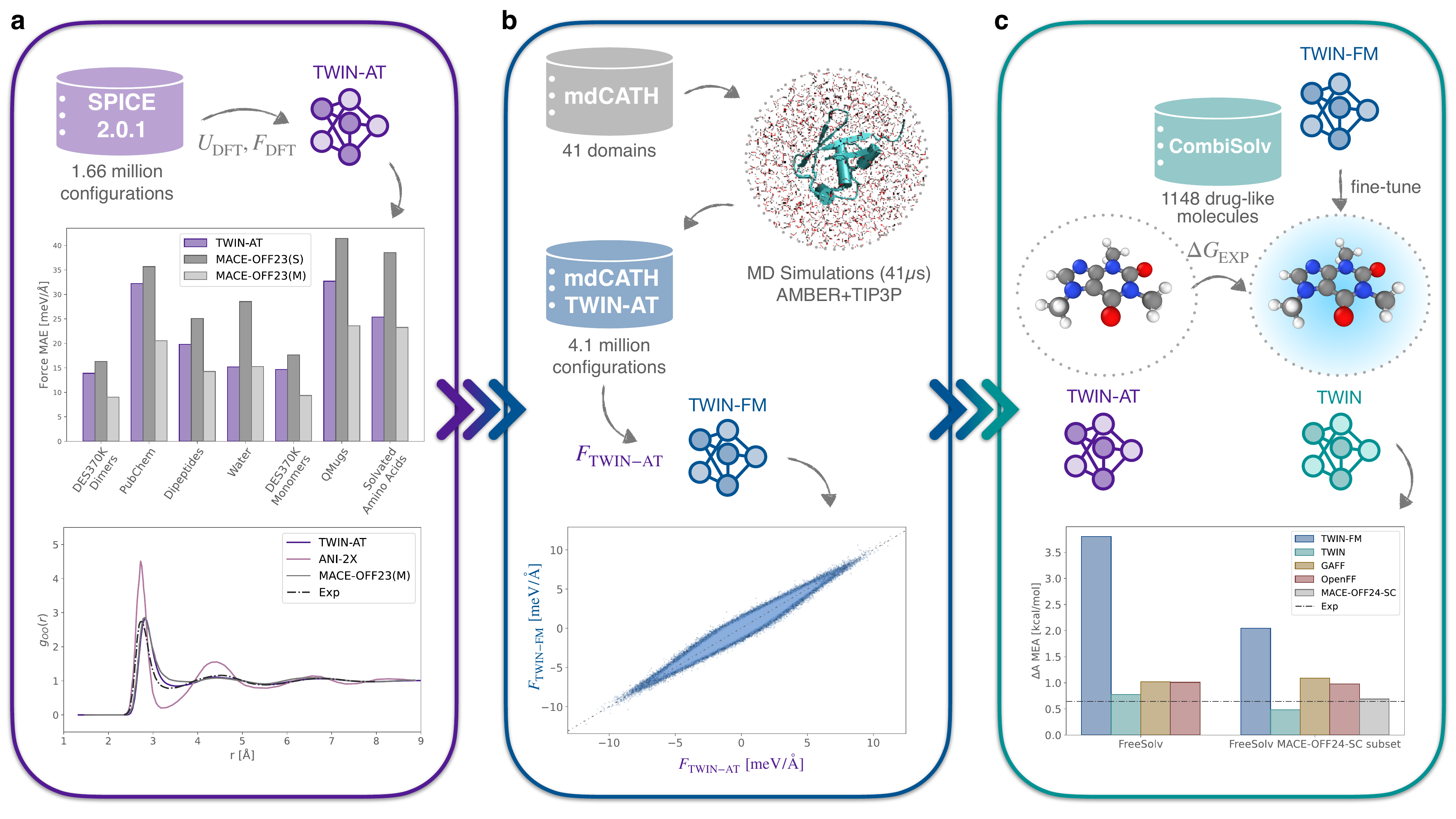} 
	\caption{\textbf{Multiscale parametrization of Transferable Water Implicit Network (TWIN) model.} (a) The TWIN-AT model, trained on the DFT-based SPICE dataset~\cite{spice2}, achieves test-set force Mean Absolute Errors (MAEs) similar to those of MACE-OFF MLP models and outperforms previous MLP models in predicting oxygen-oxygen radial distribution function ($g_{OO}$) of bulk water. The reference results are reprinted from Ref.~\cite{kovacs_2025_maceoff}. (b) The TWIN-FM model is obtained by subsequent variational force-matching training on configurations of 41 CATH domains~\cite{Sillitoe_cath_2015} generated via classical force field simulations. The parity plot shows good agreement with TWIN-AT force labels. (c) Top-down refinement of TWIN-FM against experimental solvation free energies from the FreeSolv dataset~\cite{mobley_2014_freesolv}, yielding the final TWIN MLP model. TWIN solvation free energies are drastically improved compared to TWIN-FM, approach experimental uncertainty, and outperform other classical~\cite{mobley_2014_freesolv, Karwounopoulos_freeEnergyMLP_2024} and MACE-OFF24-SC MLP model~\cite{Moore_maceoffsolvation_2026}. 
    } \label{fig:training_algorithm}
\end{figure*}

The TWIN MLP is based on the MACE architecture \cite{batatia_2022_mace} and trained using a three-stage multiscale training strategy (\autoref{fig:training_algorithm}).

In the first stage, we parametrize an atomistic (explicit solvent) reference model, TWIN-AT, using conventional bottom-up energy and force matching, as detailed in the Methods and Supplementary Information. TWIN-AT uses a compact MACE architecture with approximately $1.31 \times 10^6$ trainable parameters, which is comparable in size to MACE-OFF23(M) and approximately twice the size of MACE-OFF23(S)~\cite{kovacs_2025_maceoff}.
To reduce the computational cost of training and inference, we set the maximum spherical harmonic order to 2. 
We additionally include a very commonly used generic short-range Ziegler–Biersack–Littmark (ZBL) pair-repulsion term, which regularizes unphysical close contacts by adding a generic screened nuclear-repulsion contribution at very short interatomic distances. These design choices yield a compact and robust reference model that enables subsequent validation with extensive simulations. 
TWIN-AT is trained on 1.66 million configurations from a curated dataset comprising a preprocessed SPICE 2 subset \cite{spice, spice2} restricted to neutral configurations, together with water-cluster and QMugs subsets from the MACE-OFF training data~\cite{kovacs_2025_maceoff}. The dataset covers drug-like organic chemical space, small molecules, dimers, dipeptides, and solvated amino acids, water clusters, and QMugs molecules, with reference energies and forces computed at the $\omega$B97M-D3(BJ)/def2-TZVPPD level of theory. On the test set TWIN-AT achieves a mean absolute error of 1.5~meV/atom and 24.6~meV/\AA{} for energies and forces, respectively. 
The force errors across the SPICE-derived subsets are consistent with the performance of the MACE-OFF models, falling between those of MACE-OFF23(S) and MACE-OFF23(M) (\autoref{fig:training_algorithm}a). This level of force accuracy is reflected in the ability of TWIN-AT to capture key experimental observables with good fidelity.
In particular, an accurate description of liquid water is essential for the development of reliable implicit solvent models. TWIN-AT reproduces the radial distribution function of liquid water in close agreement with experiment, outperforming established machine-learned potentials such as ANI-2x and even MACE-OFF23(M) \cite{Devereux_ani2x_2020, Medders_FirstPrinciplesWaterPotential_2014, Soper_RDFwater_2000}.
TWIN-AT serves as a reference model for inferring solvent-mediated effective interactions via force matching and as a baseline for solvation free energy calculations. In addition, the model parameters are used to initialize the implicit solvent (CG) models. This pretraining stage introduces a physically meaningful inductive bias, as the intrinsic interactions of the solute are only weakly perturbed by solvation. Without an explicit force field scaffold for the solute, training on atomistic reference data provides less noisy supervision than direct force matching and is essential for achieving stable MD simulations.

In the second stage, solvent effects are incorporated via variational force matching, yielding the TWIN-FM model. The dataset comprises 4.1 million configurations sampled from classical explicit-solvent simulations (AMBER, TIP3P). We selected 41 protein domains from the mdCATH database~\cite{sillitoe_2014_cath,mirarchi_mdcathlargescalemddataset_2024} following the criteria of Charron et al.~\cite{charron_2023_navigating}, namely a length of 50–75 residues, a relative shape anisotropy greater than 0.04, and helix and sheet fractions below 50\%. For each domain, configurations are extracted from 1 $\mu $s of unbiased simulations and subsequently forces are labeled with TWIN-AT. This indirect strategy is adopted because direct atomistic sampling with TWIN-AT is computationally prohibitive, with an estimated cost of $\sim 17$~years per domain. The TWIN-FM parameters are optimized by minimizing the variational force-matching objective, yielding an implicit solvent model that closely reproduces the reference force labels with a mean absolute force error of 59.8~meV/\AA{} on the test set (\autoref{fig:training_algorithm}b). The vanishing effective sample size precludes reweighting at this stage. Although this introduces a formal inconsistency, the resulting bias is corrected in a subsequent stage. This bias translates to quantitatively inaccurate free energy predictions (\autoref{fig:training_algorithm}c, Supplementary Figure 5, Supplementary Figure 7). Nevertheless, the force-matching stage remains essential, as it enables generalization to larger and more flexible molecules that are not represented in existing solvation-free-energy datasets. 

In the last stage, we obtain the final TWIN model by fine-tuning it on experimental hydration free energies from the CombiSolv dataset \cite{Vermeire_combisolv_2021}. 
CombiSolv is a curated database of experimentally measured solvation free energies for solute–solvent pairs, from which we extract the aqueous subset comprising hydration free energies for 1,148 solutes covering drug-like organic chemical space.
Since solvation free energies are computed via MD simulations rather than being direct model outputs, this stage involves a top-down optimization. To this end, we employ the Solvation Free Energy Path Reweighting (ReSolv) \cite{roecken_2024_predictingsolvationFE} method that follows a free-energy path during model refinement and uses Differentiable Trajectory Reweighting (DiffTRe) \cite{thaler_2021_learning} and a hybrid approach of free-energy perturbation \cite{Zwanzig_fep_1954} and the Bennett Acceptance Ratio (BAR) method \cite{Bennett_bar_1976} to evaluate parameter updates efficiently, allowing experimental thermodynamic observables to guide the final model calibration without differentiating through complete simulation trajectories. 
TWIN predicts hydration free energies for the aqueous CombiSolv subset with a mean absolute error (MAE) of 0.76 kcal/mol (\autoref{fig:training_algorithm}c). Evaluation on FreeSolv, a commonly used hydration-free-energy benchmark that is partly represented in the experimental CombiSolv curation, yields a MAE of 0.96 kcal/mol.
This level of accuracy is close to the average experimental uncertainty ($\approx$0.6 kcal/mol) reported for FreeSolv and compares favorably with established simulation-based methods. For example, classical explicit-solvent baselines on FreeSolv subsets report MAEs of 1.02 kcal/mol (Amber/GAFF) and 1.05 kcal/mol (CHARMM/CGenFF) \cite{roecken_2024_predictingsolvationFE}. Since solvation free energy calculations are very demanding for explicit solvent MLPs, the recent MACE-OFF24-SC model was tested on a subset of 36 molecules yielding a MAE of 0.69 kcal/mol \cite{Moore_maceoffsolvation_2026}. On this subset, TWIN achieves an MAE of 0.48 kcal/mol, outperforming both MACE-OFF24-SC and classical force fields such as GAFF (1.09 kcal/mol) and OpenFF 2.1 (0.98 kcal/mol). These results demonstrate state-of-the-art performance in solvation thermodynamics across MLPs and classical force fields.

Using extensive benchmarking, we now show that this combination of training approaches, using inductive pre-training bias, bottom-up force-matching, and targeted top-down refinement against experimental thermodynamic observables, allows the TWIN model to maintain numerical stability, transferability to large systems, and quantitative accuracy.  

\subsection*{Drug‑like Molecules} 
\begin{figure*}[h]
	\centering
	\includegraphics[width=\textwidth]{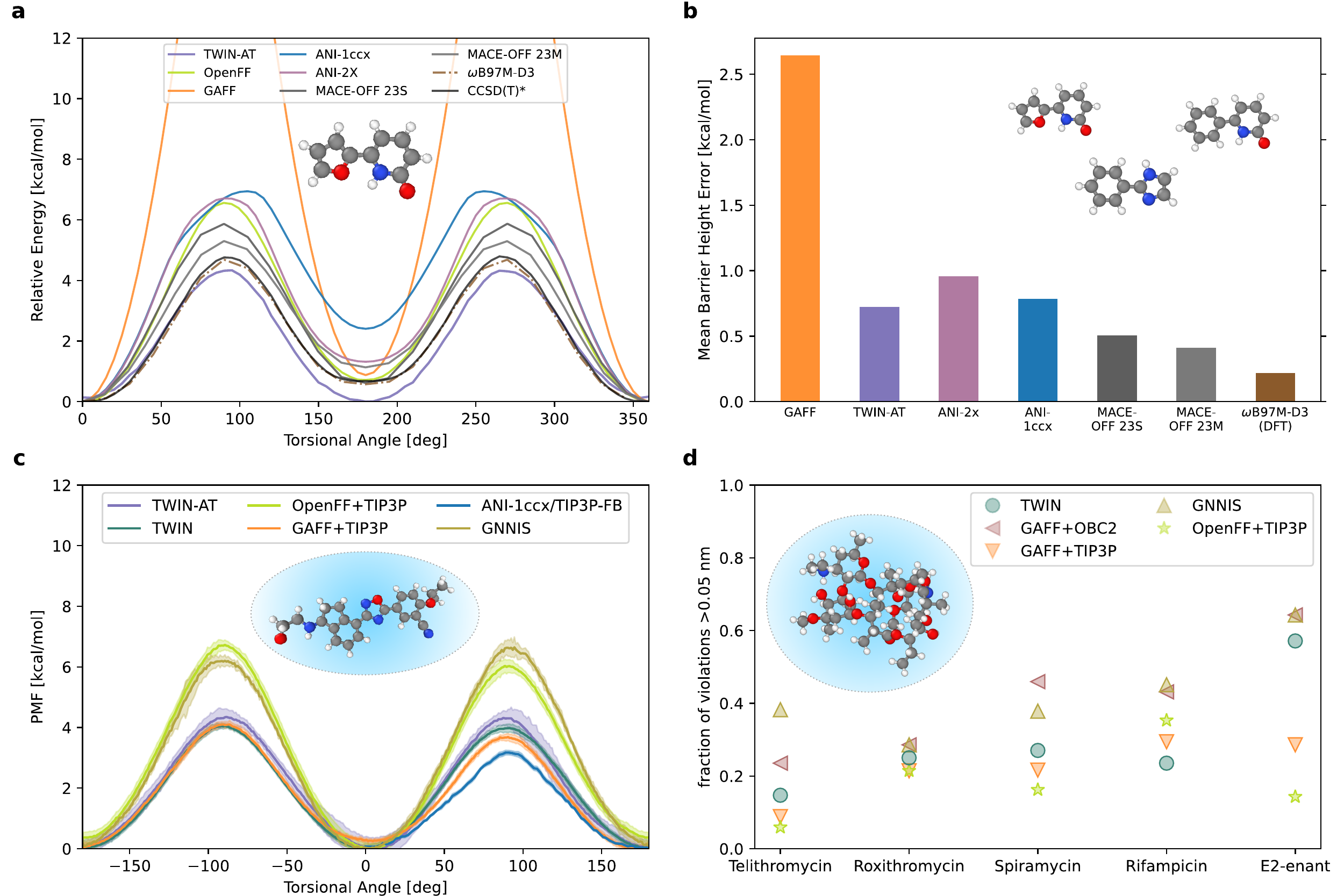} 
	\caption{\textbf{Drug-like molecules in vacuum using biaryl torsion benchmark~\cite{lahey_2020_biaryl} (a,b) and water solvent (c,d).} Torsional potential energy surface for structure 25 (a) and mean absolute deviation of rotational barrier heights relative to CCSD(T)* reference profiles (b). Data for OpenFF, GAFF, ANI-1ccx, ANI-2x, and CCSD(T)* are taken from Ref.~\cite{lahey_2020_biaryl}, while MACE-OFF models and $\omega$B97M-D3 are from Ref.~\cite{kovacs_2025_maceoff}.
		(c) Potential of mean force (PMF) for rotation around the biaryl torsion angle $\varphi$ in ozanimod, linking the isopropyl benzonitrile and diazafuran moieties. PMFs are computed for solvated ozanimod using 20~ns enhanced sampling MD simulations. The shaded region denotes one standard deviation.
		(d) Fraction of NOE violations exceeding 0.05 nm for macrocyclic compounds in aqueous solution. Results from TWIN simulations are compared to GNNIS~\cite{katzberger_2024_ageneralgraph} classical force field predictions reported by Waibl et al.\cite{noeMacrocyclicCompounds_Waibl_2024}.
    }
	\label{fig:drugs}
\end{figure*}

Drug‑like small molecules frequently contain flexible aromatic subunits whose internal conformational preferences play a major role in determining their behavior in solution. In particular, biaryl fragments introduce a torsional degree of freedom that is highly sensitive to the balance of conjugation, steric repulsion, and intramolecular electrostatic interactions  \cite{lahey_2020_biaryl}. Although these effects are already challenging to capture in the gas phase, in condensed phases, additional contributions from solute-solvent interactions can influence the relative stability of the torsional conformers and modify the effective free‑energy barriers associated with rotation.
We evaluated the accuracy of our TWIN-AT model using a biaryl torsion benchmark of 88 biaryl fragments in vacuum relevant to drug‑like chemical space \cite{lahey_2020_biaryl} (\autoref{fig:drugs}a,b). For each fragment, we compared the model‑predicted vacuum‑phase torsional potential energy surfaces with reference data obtained from coupled cluster calculations. To ensure comparability with previous studies, we restrict our analysis to 78 molecules. Across this subset, TWIN-AT achieves sub‑kcal/mol accuracy, with a mean barrier height error of 0.72 kcal/mol, demonstrating that it accurately captures the key electronic factors governing biaryl rotational energetics. This performance is comparable to that of ANI MLP models and significantly improves upon classical force fields such as GAFF (2.6 kcal/mol), which tend to overestimate torsional barriers \cite{lahey_2020_biaryl}. However, because some models exhibit substantial deviations from the high-level CCSD reference potential energy surfaces (Supplementary Fig. 3), the root-mean-square deviation (RMSD) provides a more comprehensive measure of overall accuracy (Supplementary Fig. 2a). According to this metric, TWIN-AT again performs markedly better, with an RMSD of 0.48 kcal/mol compared to 1.5 kcal/mol for GAFF.

The strong performance of atomistic MLPs, including TWIN-AT, on drug-like molecules is expected, given the composition of the SPICE 2 dataset~\cite{spice2}, which contains approximately 16,000 aryl and heteroaryl molecules. In contrast, biaryl compounds constitute a stringent test for the TWIN model. Notably, during the second stage, the model is trained exclusively on large protein fragments, whereas the third-stage FreeSolv dataset includes only 16 biaryl compounds and is dominated by small, fragment-like species, with about 95\% of the molecules having molecular weights below 300 Da. 
To assess how well TWIN recovers solvent‑mediated conformational energetics of drug-like molecules, we examine the PMF associated with rotation around the biaryl bond in ozanimod (404 Da), a therapeutic small molecule used in the treatment of multiple sclerosis. Note that ozanimod is not part of the FreeSolv dataset and thus presents a generalization test. The molecule features an isopropyl benzonitrile linked to a diazafuran ring (Supplementary Fig. 5a), forming a sterically and electronically nontrivial torsional degree of freedom \cite{ozanimod_Cohen_2016}. We used well-tempered metadynamics to enhance sampling of the torsional dihedral angle around the biaryl bond in aqueous solution (supplementary material Sec. 2,  \autoref{fig:drugs}c). 
We find that TWIN quantitatively reproduces the TWIN-AT PMF for ozanimod, with a barrier height error of 0.293 kcal/mol within the statistical uncertainty. It preserves the key features, including the positions of the minima, the barrier location, and the overall shape of the free energy profile (\autoref{fig:drugs}c). 
Comparison across the training stages (Supplementary Fig. 5 b) demonstrates that the third training stage systematically drives the PMF towards the atomistic reference (TWIN-FM barrier height error 0.764 kcal/mol). These results indicate that TWIN captures the dominant solvation effects governing this torsional degree of freedom.
The PMF predicted by TWIN agrees closely with explicit solvent MLP ANI‑1ccx results, supporting the accuracy of the learned torsional energetics. Among classical force fields, GAFF reproduces the profile well, whereas OpenFF underestimates the stability of the planar conformer, leading to barrier errors exceeding 2 kcal/mol. This discrepancy is already present in vacuum torsion scans (\autoref{fig:drugs}a), indicating that it originates from the intramolecular potential rather than solvent effects. As expected, the ML-based generalized Born model, GNNIS \cite{katzberger_2024_ageneralgraph}, closely follows the data-generating OpenFF results and inherits the model's systematic bias.

Next, we extend our analysis to macrocyclic compounds, an important class of drug‑like molecules. While their conformational freedom is restricted by ring closure, macrocyclic compounds still retain sufficient flexibility to access biologically relevant conformations \cite{macrocyclesfordrugdiscovery_Driggers_2008}. This provides macrocycles with high binding affinities while maintaining favorable properties for therapeutic development, even at molecular sizes that exceed conventional small‑molecule guidelines. Because their conformational behavior is highly dynamic and strongly influenced by the surrounding environment, accurate prediction of macrocyclic conformational ensembles remains a central challenge for computational modeling \cite{SimulationReveals_Sethio_2023}. 
Nuclear Overhauser effect (NOE) upper distance bounds obtained from NMR spectroscopy provide experimentally derived constraints on the spatial proximity of hydrogen atoms in solution and therefore serve as a rigorous benchmark for evaluating computed conformational ensembles \cite{PeptidicMacrocycles_Kamenik_2028}. To assess the accuracy of TWIN in this context, we benchmark its sampled ensembles against these NOE‑derived interproton distance bounds.
Specifically, we compute NOE‑averaged distances from 40~ns unbiased MD simulations and compare them with the corresponding experimental constraints. 
Overall, TWIN outperforms classical and ML-based implicit solvent models, GNNIS and GAFF+OBC2 (\autoref{fig:drugs}d). 
Computed conformational ensembles yield NOE‑derived hydrogen distances that are broadly consistent with the experimental values. 
A more detailed analysis of the violated distances reveals that the largest discrepancies ($>$0.1~nm) predominantly occur for hydrogen–hydrogen pairs separated by more than four bonds (Supplementary Fig. 4), suggesting that the errors are associated with non-local conformational correlations. Because the present TWIN model, based on a local message passing graph neural network MACE architecture, uses a finite local cutoff of 0.5 nm, such correlations may be only indirectly represented through the local environment and message-passing depth. This points to a plausible limitation of the underlying local architecture for flexible macrocycles. 

\subsection*{Peptides}

We consider the alanine tripeptide $\mathrm{Ala_{3}}$ as a representative peptide system and examine the models' capability to reconstruct the free energy landscape. While $\mathrm{Ala_{3}}$ itself is not part of the SPICE dataset, $\mathrm{Ala_{3}}$-like fragments are represented in two out of 41 CATH domains included in the force-matching dataset. 
This system serves as a fundamental benchmark in CG modeling, as it requires capturing a free energy surface characterized by multiple metastable conformational states. We employed enhanced-sampling MD simulations with a combined simulation time of 40 ns (see Supplementary Information). Backbone conformations of alanine peptides can be classified into a small set of mesostates in Ramachandran space, including the antiparallel $\beta$-sheet region ($a\beta$; $\Phi < -130\degree, \Psi > 120\degree$), the $\beta$-transition region ($\beta t$; $-130^\degree \leq \Phi \leq -90^\degree, \Psi > 120^\degree$), the right-handed $\alpha$-helix ($\alpha_R$; around $\Phi \approx -60\degree$, $\Psi \approx -60\degree$), and the polyproline II (pPII)-type region (pPII; $\Phi \approx -75\degree$, $\Psi \approx 145\degree$), which dominates unfolded peptide ensembles in water \cite{Zhang_alanineInWater_2020, DynamicsLysineSide_Esadze_2011}.

\begin{figure*}[h]
	\centering
	\begin{subfigure}[b]{\textwidth}
		\centering
		\includegraphics[width=0.9\textwidth]{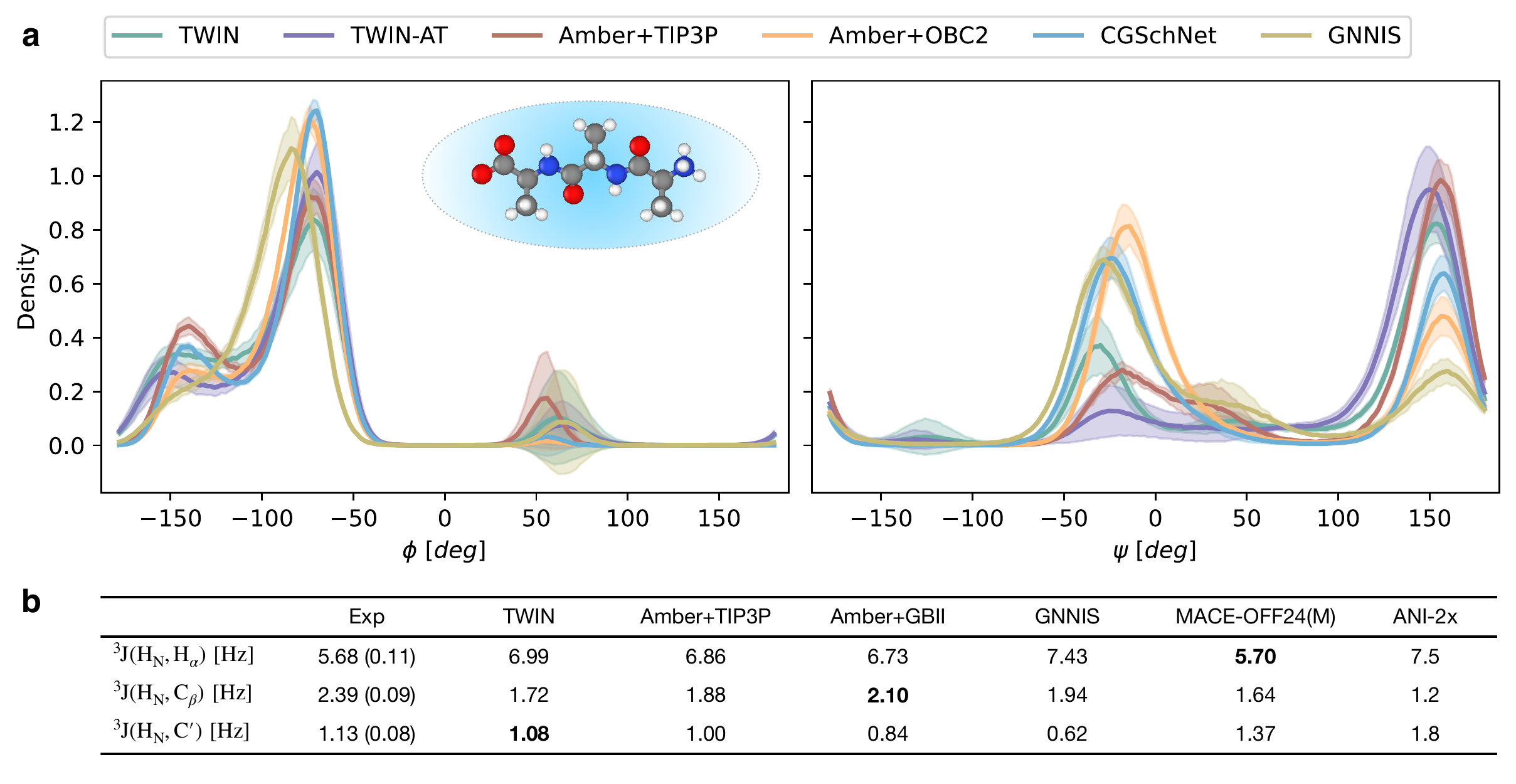}
	\end{subfigure}
	\caption{
    \textbf{Configurational analysis of $\mathrm{Ala_{3}}$ in water solvent.} (a) One‑dimensional $\phi$ and $\psi$ dihedral angle distributions of the central alanine residue. The shaded area represents the standard deviation, computed using block averaging. (b)  Three-bond scalar couplings calculated from dihedral distributions of the central residue. Values in parentheses indicate experimental uncertainties. Bold numbers indicate the closest agreement with experimental data. Experimental values are taken from Ref.~\cite{Zhang_alanineInWater_2020}, ANI-2x results from Ref.~\cite{Rosenberger_peptidesmlp_2021} and MACE-OFF24(M) data from Ref.~\cite{kovacs_2025_maceoff}.
        }\label{fig:ala3_dihedral}
\end{figure*}

The atomistic TWIN-AT model captures key qualitative features of the conformational ensemble of $\mathrm{Ala_3}$ (\autoref{fig:ala3_dihedral}a). In particular, it reproduces pPII as the dominant mesostate while maintaining a low $\alpha_R$-helical fraction ($\approx 0.06$), consistent with experimental observations \cite{Zhang_alanineInWater_2020}. The MACE-OFF24(M) model yields qualitatively similar Ramachandran distributions, suggesting a comparable description of the underlying conformational landscape (Supplementary Fig. 6). The marginal dihedral distributions of TWIN closely follow those of the explicit-solvent TWIN-AT model, with both models assigning the highest density to the pPII-associated basin. 
Comparison across training stages shows that TWIN-FM, obtained after the initial force-matching stage, already samples the conformational basins present in the explicit-solvent reference. However, it exhibits a systematic shift of the pPII basin toward smaller $\psi$ values, leading most notably to a reduced pPII population relative to TWIN-AT. The subsequent ReSolv refinement partially mitigates this shift and improves the relative thermodynamic weighting of the basins (Supplementary Fig. 7).
Models trained on classical force field data show a qualitatively different behavior. Although CGSchNet and GNNIS were trained on explicit-solvent Amber+TIP3P data, both models favor the $\alpha_R$ basin, with elevated $\alpha_R$ populations of 0.26 and 0.22, respectively, and reduced pPII sampling. This trend closely mirrors the $\alpha_R$-stabilization bias of Amber+OBC2, where $\alpha_R$ becomes the dominant mesostate ($\alpha_R \approx 0.28$, pPII $\approx 0.14$). By contrast, the explicit-solvent Amber+TIP3P reference favors pPII over $\alpha_R$ (0.33 vs. 0.08), in closer agreement with experimental observations \cite{Zhang_alanineInWater_2020}.

We further assessed the Ala3 conformational ensemble by computing the three-bond J-coupling constants from central-residue dihedral-angle distributions sampled in 20 ns unbiased MD simulations (\autoref{fig:ala3_dihedral}b) \cite{Zhang_alanineInWater_2020}. TWIN yields values in very good agreement with experiment and comparable to those of established force fields and other ML potentials, suggesting that it captures conformational distributions relevant to these NMR observables. 
Residual discrepancies are expected given the sensitivity of J-couplings to finite-sampling, Karplus parameters, and protocol choices, including the use of classical rather than path-integral MD \cite{Engel_importanceNQE_2021}. Within these limitations, the results support the use of TWIN for generating correct peptide conformational ensembles.

\subsection*{Proteins}

Having demonstrated excellent performance for peptides, we now assess the transferability of TWIN to larger proteins. We consider four benchmark systems: ubiquitin, the protein G B1 domain (GB1), cold shock protein A (CspA), and intestinal fatty acid binding protein (IFABP). These proteins were selected because extensive experimental data are available. A combination of high-resolution crystallographic structures, available as PDB entries 1UBQ, 2QMT, 1MJC, and 1IFC, and extensive NMR data define their structure and dynamics in aqueous solution. As a result, these systems are widely used as standard benchmarks for validating and fine-tuning empirical force fields \cite{huang_2013_CHARMM36nmr,ScrutinizingMolecularMechanics_Lange_2010, SystematicValidation_LindorffLarsen_2012, AreProteinForce_Beauchamp_2012}. Importantly, the selected proteins exhibit low sequence similarity, between 15 and 29 percent, relative to the data used during the second training stage, making them suitable for assessing generalization. For all systems and molecular models, we perform 10 ns unbiased MD simulations starting from the corresponding PDB structures. The proteins contain about 1,300 atoms, corresponding to 49–74 residues, and are solvated in boxes that yield systems of up to 31,500 atoms in explicit solvent. These spatiotemporal scales are beyond the practical limits of current explicit solvent MLPs, so we restrict comparisons to classical force fields. Additional computational details are provided in the Supporting Information.

\begin{figure*}[h!]
		\centering
		\includegraphics[width=\textwidth]{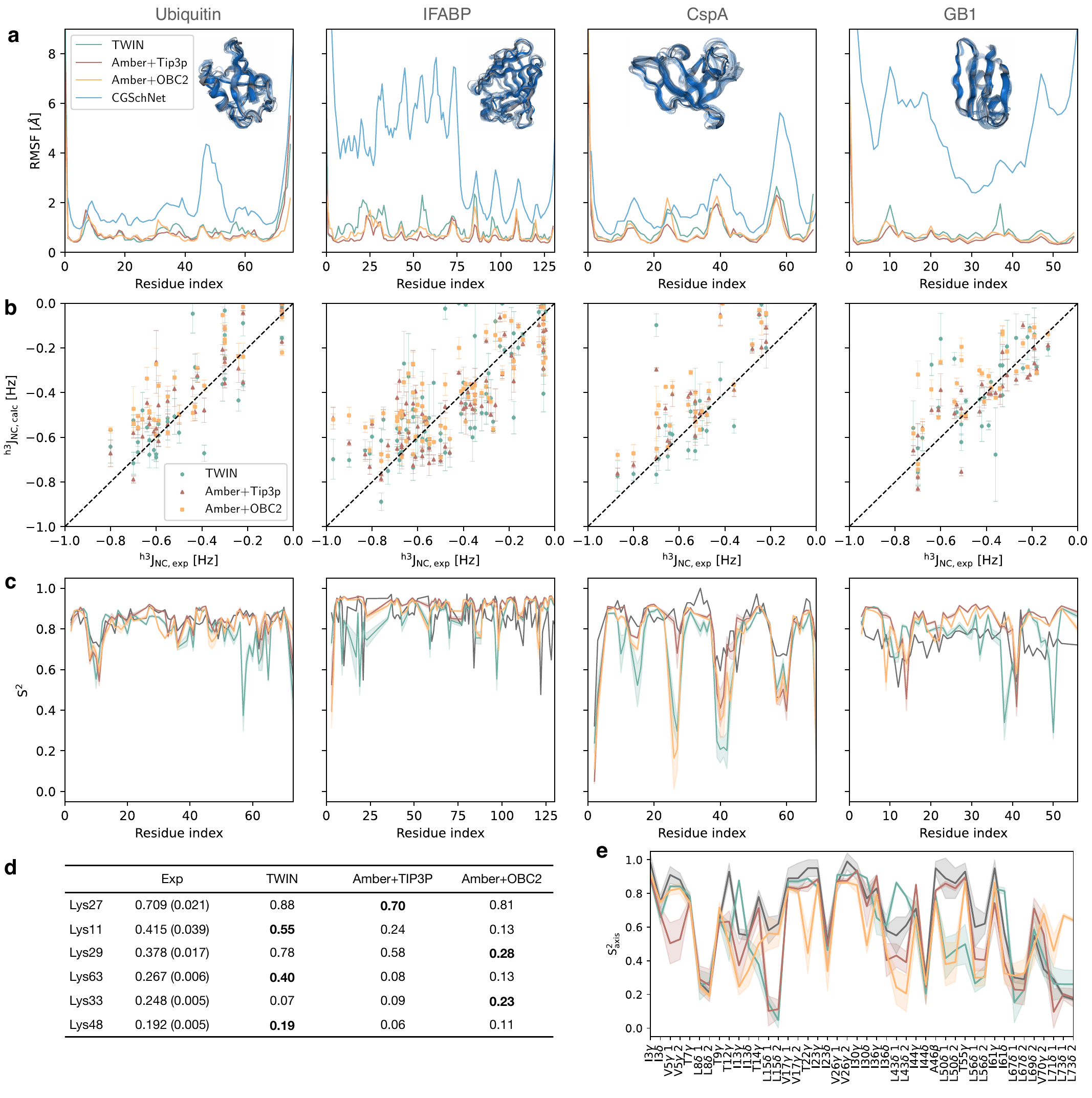}
	\caption{\textbf{Stability and flexibility of (left to right) Ubiquitin, IFABP, CspA, and GB1 proteins.}  (a) $\mathrm{C_{\alpha}}$ root mean square fluctuation (RMSF) with respect to the crystal structure. (b) Parity plots of the hydrogen bonds scalar couplings $^{h3}J_{NC}$, 
    with experimental data taken from Ref.~\cite{juranic_h3nc_1ubq_2002} (ubiquitin, IFABP), \cite{alexandrescu_h3nc_1mjc_2001} (CspA), \cite{Cornilescu_h3jnc_2qmt_1999} (GB1). The error bars represent the standard deviation, computed using block averaging. (c) Backbone N-H order parameter $\mathrm{S^2}$. The relaxation experiments values \cite{Tjandra_15NNMRrelaxation_1995, huang_2013_CHARMM36nmr} are shown in gray. (d) Relaxation order parameter $\mathrm{S^2_{axis}}$ for the lysine $\mathrm{NH_3^+}$ groups in ubiquitin \cite{esadze_j3cn_1ubq_2011}. Bold numbers indicate the closest agreement with experimental data. (e) Side-chain methyl order parameter $\mathrm{S^2_{axis}}$ 
    in ubiquitin. Experimental values are taken from Ref.~\cite{lee_s2_methyl_1ubq_1999}.  
    }\label{fig:h3jnc}
\end{figure*}

We begin by referencing simulations against X-ray crystallographic structures. We evaluate structural stability by computing the root mean square deviation (RMSD) relative to the experimental reference (Supplementary Fig. 7). TWIN maintains stable, folded conformations that remain close to the experimental structures throughout the simulations, mirroring trends observed with the Amber force field. The RMSD values are within 1–2.5 \AA{}. In contrast, CGSchNet exhibits substantially larger deviations, with RMSD values exceeding ~3–5 \AA{} and partial unfolding occurring for IFABP and GB1. The root mean square fluctuation (RMSF) analysis shows that TWIN reproduces behavior broadly comparable to classical Amber simulations, effectively distinguishing between structured and unstructured residues while showing slightly increased flexibility in less rigid regions (\autoref{fig:h3jnc}a). The inclusion of long-range electrostatics, which the current TWIN architecture lacks, has been shown to enhance rigidity in other MLPs~\cite{AUniversalAugmentationFramework_Kim_2025}, and its inclusion would likely align results more closely with classical force field predictions. On the other hand, classical force fields are known to over-stabilize ordered or compact conformations, leading to excessive helicity or overly collapsed ensembles and thus often struggle to model intrinsically disordered protein domains accurately \cite{Henriques_mdDisorderedProteins_2015, Huang_ffdisorderedproteins_2018, Rauscher_StructuralEnsemblesIDP_2015}. Unlike the other models, CGSchNet yields markedly elevated RMSF values throughout the protein, even in well-structured regions of ubiquitin and CspA. This suggests an overestimation of conformational flexibility, likely due to the greater information loss from its coarser representation \cite{Husic_cgmd_2020, Noid_cgmodelsBio_2013, Kar_AdvancesinProteinChemistry_2014}.

Next, we benchmark the models against comprehensive NMR data, providing structural information on hydrogen bonds and on backbone and side-chain flexibility. Scalar couplings across hydrogen bonds $^{h3}J_{NC}$ reflect the interaction between nitrogen and carbon nuclei mediated by $\mathrm{N-H \cdots O=C}$ hydrogen bonds \cite{Correlationbetween_Cornilescu_1999, Insightsinto_Grzesiek_2004}. Their magnitude is highly sensitive to hydrogen-bond geometry, as well as to the underlying hydrogen bonding network dynamics and cooperativity within proteins \cite{huang_2013_CHARMM36nmr, StructuralDependencies_Barfield_2002, Keystabilizingelements_Nisius_2012}. Across all proteins, TWIN shows good agreement with experimental hydrogen-bond scalar couplings (\autoref{fig:h3jnc}b), with performance that is broadly comparable to established classical force fields (Supplementary Table~4). While TWIN predictions exhibit a somewhat larger spread and uncertainty than explicit solvent AMBER simulations, the deviations remain modest across all systems. Notably, for ubiquitin, TWIN achieves particularly strong agreement, approaching the accuracy of AMBER+TIP3P and outperforming the implicit solvent Amber+OBC2 model. 
Perfect agreement with experiment is not expected, as classical MD neglects nuclear quantum effects that can influence hydrogen bond geometries and thus $^{h3}J_{NC}$ values, particularly at short donor–acceptor distances below 2.8~\AA{} \cite{Zhou_shorthbonds_2019, Ogata_nqe_2013}. Differences below about 0.1~Hz therefore fall within the combined range of experimental uncertainty and methodological limitations \cite{alexandrescu_h3nc_1mjc_2001, Cornilescu_h3jnc_2qmt_1999, Cornilescu_IdentificationHBN_1999}. 

We proceed with backbone dynamics analysis and compare simulated $\mathrm{N-H}$ order parameters $\mathrm{S^2}$ with experimental values (\autoref{fig:h3jnc}c). The calculated order parameters show overall good agreement with experiment across all proteins, indicating that TWIN reproduces the balance between rigid and flexible regions at a global level. At the same time, the TWIN trajectories tend to predict increased backbone mobility relative to AMBER simulations (Supplementary Fig.~9). This effect is most pronounced in regions with intrinsically low order parameters, where $\mathrm{S^2}$ values are systematically underestimated, suggesting that highly flexible segments are described as overly dynamic. In contrast, residues with high experimental stiffness are well reproduced, with $\mathrm{S^2}$ values in these regions comparable to those obtained from the AMBER trajectories.

Next, we examine the side chain structural properties. Lysine side chains are important contributors to protein function and stability, as they participate in hydrogen bonds, form salt bridges with acidic groups, and thereby influence the strength and specificity of intermolecular interactions \cite{esadze_j3cn_1ubq_2011}. Their $\mathrm{NH3^+}$ groups are highly dynamic, and these dynamics play crucial roles in the various functions performed by ubiquitin, including its involvement in enzymatic and protein–protein interactions \cite{SignatureofMobile_Zandarashvili_2011}. The $\mathrm{^{15}N}$ relaxation order parameters $\mathrm{S^2_{axis}}$ of lysine $\mathrm{NH_3^+}$ groups characterize the degree of motional restriction around the $\mathrm{C_{\gamma}N_{\zeta}}$ axis. 
TWIN captures the experimentally observed heterogeneity of lysine side-chain dynamics, correctly identifying Lys27 as the most restricted residue while reproducing the high mobility of Lys48 (\autoref{fig:h3jnc}d). 
For the remaining lysines, the predicted order parameters reflect pronounced site-specific differences in the $\mathrm{NH_3^+}$ motions. 
The Amber+TIP3P and Amber+OBC2 results further illustrate that the lysine $\mathrm{S^2_{axis}}$ values depend sensitively on the solvent representation. These results indicate that TWIN provides a balanced description of lysine side-chain dynamics, capturing both highly mobile and more restricted regimes. 

To further probe side-chain dynamics, we evaluated inter-residue three-bond scalar couplings $\mathrm{^3J_{C_{\gamma}N_{\zeta}}}$ in lysine side chains from MD trajectories of ubiquitin (Supplementary Fig.~10). TWIN samples the major $\chi_4$ rotameric states observed in the Amber reference simulations, although the relative populations differ for several lysine residues. The resulting scalar couplings show deviations from experiment for some sites, reflecting the sensitivity of this observable to the underlying rotamer distribution. Since $\mathrm{^3J_{C_{\gamma}N_{\zeta}}}$ depends directly on the $\chi_4$ dihedral angle, differences in the predicted couplings can be attributed to differences in the lysine side-chain dihedral distributions. This is consistent with previous observations by Huang et al.~\cite{esadze_j3cn_1ubq_2011} that inter-residue lysine couplings are highly sensitive to small changes in rotamer populations. Finite sampling of the slowly interconverting lysine rotamers may further contribute to the remaining deviations.

Extending our analysis beyond lysine $\mathrm{NH_3^+}$ groups, we also examine side-chain methyl ($\mathrm{CH_3}$) groups in ubiquitin (\autoref{fig:h3jnc}e). The AMBER simulations reproduce the experimental measurements of side-chain methyl order parameter $\mathrm{S^2_{axis}}$ well, whereas TWIN generally yields smaller order parameters, indicating increased mobility of the methyl side chains. This behavior mirrors the trends observed for backbone dynamics, where TWIN similarly predicts enhanced flexibility relative to AMBER. The consistent underestimation of $\mathrm{S^2}$ across both backbone and side-chain observables suggests a systematic tendency toward more dynamic conformational sampling in TWIN. We attribute this discrepancy to the predominantly local nature of the TWIN model and anticipate that incorporating explicit long-range interactions would improve agreement with experiment \cite{AUniversalAugmentationFramework_Kim_2025}.

Taken together, the extensive structural analysis across four proteins establishes TWIN as a reliable approach for modeling protein native states. It greatly improves the stability relative to previous CG MLPs and reproduces NMR predictions at a level competitive with established explicit solvent force fields.

\subsection*{Protein-Ligand Interactions}
An accurate description of protein–ligand intermolecular interactions is essential for molecular modeling, as these interactions directly determine binding energetics and ligand affinity, which are of particular interest in drug design. Since such interactions arise from a combination of local pairwise contributions and nonadditive many-body effects, a systematic evaluation of a model should ideally disentangle these contributions across different levels of complexity.

We begin our evaluation on pairwise protein–ligand interactions using the PLF547 dataset \cite{Kristian_pla15_2020}, which consists of complexes formed between ligands and individual protein fragments extracted from experimentally resolved structures. In this dataset, protein sites are decomposed into fragments by cutting the backbone and side chains into capped segments that preserve the local chemical environment of the binding site. The resulting ligand–fragment complexes isolate individual interaction contributions while retaining realistic geometries and chemical diversity. Interaction-energy labels are obtained from reference calculations based on a composite MP2-F12 + $\Delta$DLPNO-CCSD(T) protocol. 
Restricting the analysis to neutral–neutral systems to remain consistent with the architecture of TWIN-AT, we obtain a MAE of 0.48 kcal/mol (RMSE = 0.65 kcal/mol). This level of accuracy is consistent with established semiempirical quantum mechanical methods (RMSE of 0.5 - 1.3 kcal/mol) and other foundational MLPs like MACE-OMOL or UMA (MAE of 0.6 - 1.1 kcal/mol) \cite{Kristian_pla15_2020, batatia_macepolar_2026}. 

\begin{figure*}[h]
	\centering
	\includegraphics[width=\textwidth]{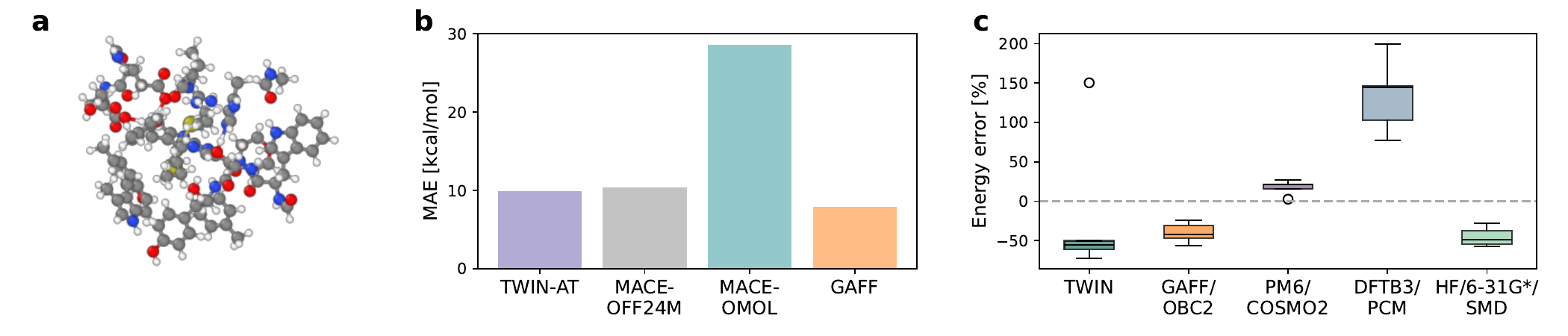} 
	\caption{\textbf{Protein-ligand interactions.} (a) An example complex from the PLA15 benchmark \cite{Kristian_pla15_2020}. (b) Mean absolute errors (MAE) of protein-ligand interaction energies in vacuum for neutral ligand complexes in PLA15. (c) Distribution of errors for interaction energy changes of protein-ligand complexation upon solvation, relative to COSMO-RS predictions. Values for semiempirical quantum mechanical methods PM6/COSMO2 and DFTB3/PCM, as well as HF/6-31G*/SMD, are taken from Ref.~\cite{Kristian_pla15_2020}. 
    }
	\label{fig:pla15}
\end{figure*} 

Next, we extend our analysis to complete active-site models contained in the PLA15 dataset, which reconstruct ligand interactions within their full environment and thus incorporate many-body nonadditive contributions. We perform our analysis on a subset of PLA15 containing complexes with neutral ligands. 
In vacuum (\autoref{fig:pla15}b), TWIN-AT yields interaction energies with accuracy comparable to established force fields and MLPs, with errors on the same order as MACE-OFF24M, MACE-OMOL, and GAFF.
While modeling protein–ligand interactions in vacuum provides a useful baseline, practical applications require an accurate description of these interactions in aqueous environments \cite{Kristian_pla15_2020}. To assess the performance of TWIN, we evaluate the change in interaction energies upon solvation (\autoref{fig:pla15}c) and compare it to reference values obtained with the COSMO-RS solvent model. 
TWIN systematically overestimates the magnitude of the interaction energy change, indicating a residual bias in the learned solvent response. Nevertheless, its overall accuracy is still better than that of the implicit solvent semi-empirical quantum mechanical model DFTB3/PCM and comparable to the classical generalized Born implicit solvent and HF/6-31G*/SMD semi-empirical model. These results suggest that the solvent effects introduced during training stages two and three are captured at a qualitatively correct level, enabling the model to account for solvent-mediated interactions and supporting its application to protein–ligand systems in solution.

\subsection*{Computational Performance}

The primary motivation for developing an implicit solvent model is to reduce computational cost. To quantify the efficiency gains offered by TWIN relative to the explicit solvent TWIN-AT, we compare the computational performance of MD simulations for $\mathrm{Ala_3}$ and ubiquitin benchmark systems used in this study. The simulations are conducted on an NVIDIA A100 80GB PCIe GPU and employ LAMMPS \cite{Thompson_lammps_2022} with chemtrain-deploy \cite{Fuchs_chemtrainDeploy_2025} plugin utilizing OpenEquivariance~\cite{Bharadwaj_openequivariance_2025} and a timestep of 0.5~fs. Explicit solvent TWIN simulations were initialized from water configurations equilibrated at a density of $1~\mathrm{g/cm^3}$. Depending on system size, TWIN achieves one to two orders of magnitude speedup per MD timestep compared to the explicit solvent approach (Supplementary Table 5). 
The speedup arises from the reduced number of particles in the implicit solvent representation, for which near-linear scaling of the computational cost is expected. In practice, however, performance is also governed by GPU utilization, which remains suboptimal for small systems (e.g., Ala3). Consequently, even greater speedups over all-atom representations are anticipated for larger or more dilute systems.

For context, we additionally report the computational performance of GNNIS and CGSchNet using their native software on the same hardware. We find that timestep evaluation remains computationally more demanding for these models, including CGSchNet with a reduced number of degrees of freedom, suggesting inefficiencies in the underlying implementations. 
We also report the computational cost of the classical AMBER ff99SB‑ILDN force field with TIP3P water. These simulations were carried out in OpenMM \cite{Eastman_OpenMM_2024} with GPU acceleration on identical hardware under the same settings as in the LAMMPS simulations. 
Compared to TWIN, simulations with classical force fields yield up to a 25-fold increase in simulation speed. 
However, library support and software implementations for MLP-based simulations remain at an early stage of development, and ongoing efforts such as OpenEquivariance are expected to further close this performance gap. Moreover, timestep performance does not account for the speedup from averaging out solvent degrees of freedom, which smooths the effective energy landscape and enhances sampling efficiency relative to explicit-solvent simulations. Based on previous studies, we estimate that an additional one to two orders of magnitude could be gained due to this effect, effectively making the TWIN model both cheaper and overall more accurate than explicit solvent classical force fields~\cite{Anandakrishnan_speedofconformations_2015, katzberger_2024_ageneralgraph, charron_2023_navigating}.

\section*{Discussion}\label{sec:discussion}

In this work, we presented TWIN, a transferable implicit solvent MLP for solvated proteins and drug-like molecules in aqueous environments. We benchmarked TWIN on peptides, proteins, drug-like molecules, and protein-ligand complexes outside the training dataset, demonstrating transferability across chemical space and molecular size. For small systems, including the $\mathrm{Ala_3}$ peptide and the drug molecule ozanimod, TWIN accurately reproduced the free energy landscapes obtained from explicit solvent TWIN-AT simulations, highlighting its ability to capture essential thermodynamic features across chemically distinct systems with near ab initio accuracy. At the same time, TWIN reduced computational cost by up to two orders of magnitude compared with explicit solvent simulations, enabling efficient exploration of larger systems that require extensive sampling. For protein systems, where long-timescale explicit-solvent MLP simulations remain computationally prohibitive, validation was carried out against structural and NMR experimental data. TWIN reproduced a broad range of experimental observables and surpassed established biomolecular force fields for selected observables, even though modern protein force fields represent highly optimized baselines, having undergone extensive refinement and validation against the same structural and NMR benchmarks \cite{Beauchamp_ffnmr_2012, Koes_ffnmrshifts_2017, Cavender_structureff_2025}.

The accuracy and transferability of TWIN results from the combination of a sequential bottom-up and top-down training strategy with a fully data-driven model design. In contrast to approaches trained on atomistic force field data, TWIN is trained on ab initio and experimental data. In the first stage, the atomistic reference model TWIN-AT is parameterized, providing a strong physical bias from essentially noiseless data and enabling efficient learning of fundamental interactions. 
The second stage extends the model’s applicability beyond small solvation-free-energy datasets by incorporating folded proteins and larger organic molecules from the CATH dataset, thereby enhancing transferability. Without this step, training the implicit solvent model would require solvation free energies for molecules significantly larger than those available in datasets such as CombiSolv. 
While larger resources may mitigate this limitation, they are unlikely to fully resolve it. 
Nevertheless, this intermediate stage is not formally exact for CG force matching, as the configurations are neither generated by the same model as the labels nor explicitly reweighted. Consequently, the final refinement against experimental data is essential for achieving quantitative agreement with explicit-solvent references, as demonstrated for $\mathrm{Ala_3}$ and ozanimod.

The data-driven approach further distinguishes TWIN from previous hybrid approaches. Unlike models that rely on classical force field descriptions of the solute or system-specific prior potentials, TWIN avoids imposing predefined intra-solute bonded and nonbonded interactions. This design eliminates the need for system-specific prior parameterization and supports generalization across a broad chemical space, including both proteins and small organic molecules, whereas previous ML-based CG models have typically targeted one of these molecular classes \cite{katzberger_2024_ageneralgraph, majewski_machine_2023, Wang_mlcgff_2019}. 
Hybrid architectures are often limited by the domain of applicability of their prior models. For example, protein-specific force field priors can restrict transferability beyond natural amino acids \cite{Wang_mlcgff_2019, charron_2023_navigating}. Although automated CG parameterization tools have advanced considerably, robust parameterization of arbitrary molecules remains challenging \cite{bereau2015automartini, kelidou_automated_2025}. By avoiding prior parameterization altogether, TWIN addresses this limitation while maintaining numerical stability over long simulations.

The presented workflow is general and can be extended to other solvents and MLP architectures. 
In particular, incorporating explicit long-range corrections may improve protein structural properties, distance-dependent NMR observables such as NOE violations, and protein-drug interactions \cite{smith_approaching_2019, Chen_demlpccsd_2023, Daru_ccsdmd_2022}. Generalizability could also be expanded by complementing the mdCATH dataset \cite{mirarchi_mdcathlargescalemddataset_2024} with unfolded structures, for example, by generating unfolding trajectories from folded CATH domains. Accordingly, protein folding was not considered as a target application in this work. The present benchmarks focused on equilibrium sampling around folded states, whereas de novo folding requires sampling long-timescale transition pathways and is increasingly addressed through complementary structure-prediction and generative modeling frameworks \cite{jumper_alphafold_2021, watson_novo_2023}. 
In addition to increasing the configurational diversity of the training data, the workflow could also benefit from improvements in the underlying explicit solvent MLPs.
More accurate ab initio datasets, included via multifidelity \cite{Kim_Multifidelity_2025}, transfer learning \cite{smith_approaching_2019, Zaverkin_Transferlearning_2023} or multi-head outputs \cite{Zubatyuk_multitask_2019, Batatia_foundationalmlp_2025}, will increase the accuracy of explicit solvent MLPs which are expected to cascade to implicit solvent MLPs via the presented multiscale approach. 

Overall, TWIN provides a robust and extensible framework for implicit solvent modeling, combining computational efficiency with molecular accuracy for biomolecular systems involving proteins and drug-like molecules. We expect this approach to play an important role in future biomolecular simulations as well as beyond conventional MD, for example, by providing energy labels for CG Boltzmann generators and other generative models \cite{chen_cgbg_2026, noe_bg_2019}.

\section*{Methods}\label{sec:methods}

\subsection*{Model training and validation}

We employ an implementation of the semi local message passing graph neural network MACE \cite{batatia_2022_mace} in JAX. The multistage training of the implicit water ML potential was carried out using the chemtrain framework, which provides efficient algorithms for bottom-up and top-down training \cite{fuchs_2025_chemtrain}. The hyperparameters are listed in Supplementary Table 1.

\subsubsection*{TWIN-AT model}
In the first step, we parameterize an MLP for molecules in an explicit solvent environment. The potential model takes the molecule's configurational state $S$ as input and predicts its potential energy, expressed by $U_{\text{AT}} = U(S;\theta_{\text{AT}})$. Forces on atoms are calculated from the gradient of the potential energy with respect to the atom's position. During training, we optimize the model parameters $\theta_{\text{AT}}$ in such a way that its predictions align with the reference energies $U_{\text{DFT}}$ and forces $F_{\text{DFT}}$ from the ab initio database, i.e employing a bottom-up training approach following \autoref{eqn:bottom-up_loss}.

In the first stage of our procedure, we trained an explicit solvent ML potential employing a force-matching approach. The reference dataset provides atomic configurations and corresponding forces $\vb*{f}^{ref}$ and energies $U^{\text{ref}}$. The parameters of the ML potential, $\theta$, are adjusted via backpropagation to match reference values. The respective loss function for a set of N samples is given by
\begin{equation} \label{eqn:bottom-up_loss}
	\mathcal{L}(\theta) = \frac{1}{N} \sum_{i=1}^{N} \left( \omega_U \left( U^{\theta}(\vb*{r}_i, Z_i) - U^{\text{ref}}_i \right)^2 +\frac{ \omega_F}{N_{\text{atom}, i}} \sum_{\alpha=1}^{N_{\text{atom}, i}} \norm{\vb*{f}^{\theta}_{i, \alpha}(\vb*{r}_i, Z_i) - \vb*{f}^{\text{ref}}_{i, \alpha} }_2^2  \right)
\end{equation}
where $U^{\theta}(\vb*{r}_i, Z_i)$ is the ML potential's energy prediction of the $i$-th atomic configuration in the batch and $\vb*{f}^{\theta}_{i, \alpha}(\vb*{r}_i, Z_i)$ the corresponding predicted force vector on atom $\alpha$. The relative weights of the force and energy contribution were set to $\omega_F = 1e3$ and $\omega_U = 40$ in the first training stage and changed to $\omega_F = 10$ and $\omega_U = 1e3$ for refinement. Further hyperparameters of the bottom-up training are listed in Supplementary Table 2.

Ab initio reference data for bottom-up training of the explicit-solvent ML potential were obtained from a curated dataset comprising a preprocessed subset of SPICE v2.0.1, supplemented with water-cluster and QMugs subsets from the MACE-OFF training data \cite{kovacs_2025_maceoff}. The resulting dataset comprises approximately 1.66 million configurations spanning drug-like organic chemical space, including small molecules, dimers, dipeptides, solvated amino acids, water clusters, and QMugs compounds, and containing the elements {H, C, N, O, F, P, S, Cl, Br, I, K, Li, Na}. Structures containing boron or silicon were excluded.  
Unlike MACE-OFF \cite{kovacs_2025_maceoff}, which excluded charged species entirely, our dataset retains ions and charged molecular fragments within configurations with non-zero net charge. 
This enables the model to learn interactions involving ionic and charged solute species that are commonly encountered under physiological conditions and are known to influence free-energy calculations in molecular dynamics simulations \cite{donnini_2005_incorporatingtheeffectofionicstrength}. The complete dataset was randomly divided into training (70\%), validation (10\%) and test (20\%) sets. The training was based on DFT forces and formation energies, as these do not contain any species-dependent energy shift.

\subsubsection*{TWIN-FM model and mdCATH TWIN-AT dataset}

The bottom-up training of the implicit water model is carried out by force-matching on an implicit water dataset. To this end, a total of 41 domains were selected from the CATH database \cite{sillitoe_2014_cath}. The selection criteria follow the considerations by Charron et al. \cite{charron_2023_navigating}. The selected structures are characterized by containing 50 to 75 residues, a relative shape anisotropy greater than 0.04, and a helix and sheet fraction smaller than 50 \%. A list of the selected domains is provided in Supplementary Table 4.

For all CATH domains, initial structure and topology were first extracted from the mdCATH dataset \cite{mirarchi_mdcathlargescalemddataset_2024}. The dataset contains structures whose charge states, proton placement, and hydrogen bond network optimization have been adjusted to pH 7. Each domain was then solvated in GROMACS and neutralized by adding $\text{Na}^+ $ and $\text{Cl}^-$ ions. The systems were then equilibrated in OpenMM using AMBER ff-99SB-ILDN force field with TIP3P water model for 50~ps and subsequently simulated for 1~$\mu$s using 0.5~fs timesteps, with configurations sampled every 10~ps. 
We used the Langevin thermostat with a friction coefficient of 0.1/ps to maintain a temperature of 298.15~K during simulations. Long-range electrostatic interactions were modeled using the PME algorithm with a nonbonded interactions cutoff set to 1~nm.
The simulations yielded 4.1 million protein fragment configurations in explicit water. For further use as a force matching data set, the forces acting on the atoms of the solute were calculated based on the explicit solvent model, and the solvent (water, ions) was removed. 

In bottom-up training on the implicit solvent dataset, we initially only trained the linear layers of the message passing filters of the MACE model. This ensured that the predicted solvation free energy values did not deviate too greatly from the target values. An overview of further settings and chosen hyperparameters for bottom-up training is provided in Supplementary Table 2. The complete dataset was randomly divided into training (70\%), validation (10\%), and test (20\%) sets. 

\subsubsection*{TWIN model}

For the top-down learning of the solvation free energies, Open Babel was used for the generation of 3d topologies of small organic molecules from the experimental CombiSolv dataset \cite{Vermeire_combisolv_2021} based on SMILES strings. This step also included energy minimization using the MMFF94 force field \cite{Halgren_mmff_1996, Halgren_mmff2_1996, Halgren_mmff3_1996, Halgren_mmff4_1996, Halgren_mmff5_1996} and a comprehensive conformer search. In preparation for the training, an equilibration for 50~ps was performed first, followed by a production run over 550~ps. The simulations were performed in an NVT ensemble at 298.15~K. A GSD Langevin thermostat with a friction coefficient of 50/ps was used. Using hydrogen mass repartitioning, the hydrogen atom mass was increased by a factor of 2, allowing the simulation time step to be set to 1~fs. After the simulations, stability was verified by checking whether each molecule remained a single, connected graph in the last frame of the trajectory. Unstable systems were excluded from further training, resulting in a set of 1148 trajectories. For the calculation of new trajectories during training, starting from the last frame of the previous trajectory, a 50~ps equilibration followed by a 550~ps production run was conducted. As the threshold for the recalculation of the trajectories, the effective sample size threshold was set to $\bar{N}_{\text{eff}} = 0.9$. Further hyperparameters of the top-down training of the implicit solvent model are listed in Supplementary Table 2. 

Based on the atomistic reference model $U_{\text{AT}}$ (TWIN-AT), the ML potential $U_{\text{sol}} = U(S, \theta_{\text{sol}})$ (TWIN) describing molecular interactions in an explicit aqueous environment is derived gradually. We begin with bottom-up force matching on an implicit solvent dataset containing configurations of molecules in an implicit aqueous environment. The model trained in this way provides a physically meaningful starting point for further refinement. 

After bottom-up training on the implicit solvent dataset, the implicit water ML potential (TWIN-FM) is corrected based on experimental data using a top-down learning approach following the ReSolv framework \cite{roecken_2024_predictingsolvationFE}, building on DiffTRe \cite{thaler_2021_learning}. The model parameters are optimized to reproduce experimental solvation free energies by minimizing
\begin{equation}
    \mathcal{L}(\theta) = \left(\Delta A(\theta) - \Delta A_{\mathrm{exp}}\right)^2,
\end{equation}
where $\Delta A(\theta)$ denotes the free-energy difference between reference vacuum and solvent ensembles. Since $\Delta A$ depends on molecular simulations, we employ DiffTRe \cite{thaler_2021_learning} to obtain gradients without backpropagation through the dynamics. Ensemble averages under $U_\theta$ are estimated from configurations generated with a reference potential $U_{\hat{\theta}}$,
\begin{equation}
    \langle O \rangle_\theta \approx \sum_i w_i\, O(S_i), \quad
    w_i \propto \exp[-\beta (U_\theta(S_i) - U_{\hat{\theta}}(S_i))],
\end{equation}
which yields a differentiable mapping from $\theta$ to observables. The validity of reweighting is monitored via the effective sample size $N_{\mathrm{eff}}$, and trajectories are regenerated when the overlap becomes insufficient.
Following ReSolv, the optimization path itself defines the free-energy integration pathway. 
Incremental free-energy differences between successive parameter states are computed via free-energy perturbation \cite{Zwanzig_fep_1954}, with the BAR method \cite{Bennett_freeEnergy_1976} applied whenever trajectories are recomputed, and subsequently accumulated along the training trajectory,
\begin{equation}
    \Delta A = \sum_i \Delta A_i, \qquad
    \Delta A_{\hat{\theta} \rightarrow \theta} = -\frac{1}{\beta} \ln \left\langle
    e^{-\beta (U_\theta - U_{\hat{\theta}})} \right\rangle_{\hat{\theta}}.
\end{equation}
This approach enables efficient top-down learning of free energies while avoiding differentiation through the simulation. Final solvation free energies are obtained using the BAR estimator between vacuum and solvent ensembles.

\backmatter

\bmhead{Supplementary information}

Additional details on model training, simulation protocols, benchmarks, and supporting analyses are provided in the Supplementary Information.

\bmhead{Data availability}

The simulation data required to reproduce the results presented in this manuscript will be made publicly available upon article publication. 

\bmhead{Code availability}

The code required to reproduce the results presented in this manuscript will be made publicly available upon publication of this article.

\bmhead{Acknowledgements}

The authors gratefully acknowledge the Gauss Centre for Supercomputing e.V. (www.gauss-centre.eu) for funding this project by providing computing time on the GCS Supercomputer JUWELS \cite{alvarez_juwels_2021} at Jülich Supercomputing Centre (JSC).

\bmhead{Funding} 

Funded by the European Union. Views and opinions expressed are however those of the author(s) only and do not necessarily reflect those of the European Union or the European Research Council Executive Agency. Neither the European Union nor the granting authority can be held responsible for them. This work was funded by the ERC StG (SupraModel, 101077842).

\bmhead{Author contributions} 

J.E. and J.Z. conceived the study and developed the overall computational strategy. J.E. developed and implemented the TWIN model, curated and prepared the training and validation datasets, performed the model training, molecular simulations, benchmarking, data analysis, and visualization. J.Z. supervised the project, acquired funding and computational resources, and contributed to the development of the methodology. J.E. wrote the original manuscript draft with input from J.Z. Both authors analyzed and interpreted the results, reviewed and edited the manuscript.

\bmhead{Competing interests} 

The authors declare no competing interests.

\bibliography{sn-bibliography}


\begin{thebibliography}{126}
\ifx \bisbn   \undefined \def \bisbn  #1{ISBN #1}\fi
\ifx \binits  \undefined \def \binits#1{#1}\fi
\ifx \bauthor  \undefined \def \bauthor#1{#1}\fi
\ifx \batitle  \undefined \def \batitle#1{#1}\fi
\ifx \bjtitle  \undefined \def \bjtitle#1{#1}\fi
\ifx \bvolume  \undefined \def \bvolume#1{\textbf{#1}}\fi
\ifx \byear  \undefined \def \byear#1{#1}\fi
\ifx \bissue  \undefined \def \bissue#1{#1}\fi
\ifx \bfpage  \undefined \def \bfpage#1{#1}\fi
\ifx \blpage  \undefined \def \blpage #1{#1}\fi
\ifx \burl  \undefined \def \burl#1{\textsf{#1}}\fi
\ifx \doiurl  \undefined \def \doiurl#1{\url{https://doi.org/#1}}\fi
\ifx \betal  \undefined \def \betal{\textit{et al.}}\fi
\ifx \binstitute  \undefined \def \binstitute#1{#1}\fi
\ifx \binstitutionaled  \undefined \def \binstitutionaled#1{#1}\fi
\ifx \bctitle  \undefined \def \bctitle#1{#1}\fi
\ifx \beditor  \undefined \def \beditor#1{#1}\fi
\ifx \bpublisher  \undefined \def \bpublisher#1{#1}\fi
\ifx \bbtitle  \undefined \def \bbtitle#1{#1}\fi
\ifx \bedition  \undefined \def \bedition#1{#1}\fi
\ifx \bseriesno  \undefined \def \bseriesno#1{#1}\fi
\ifx \blocation  \undefined \def \blocation#1{#1}\fi
\ifx \bsertitle  \undefined \def \bsertitle#1{#1}\fi
\ifx \bsnm \undefined \def \bsnm#1{#1}\fi
\ifx \bsuffix \undefined \def \bsuffix#1{#1}\fi
\ifx \bparticle \undefined \def \bparticle#1{#1}\fi
\ifx \barticle \undefined \def \barticle#1{#1}\fi
\bibcommenthead
\ifx \bconfdate \undefined \def \bconfdate #1{#1}\fi
\ifx \botherref \undefined \def \botherref #1{#1}\fi
\ifx \url \undefined \def \url#1{\textsf{#1}}\fi
\ifx \bchapter \undefined \def \bchapter#1{#1}\fi
\ifx \bbook \undefined \def \bbook#1{#1}\fi
\ifx \bcomment \undefined \def \bcomment#1{#1}\fi
\ifx \oauthor \undefined \def \oauthor#1{#1}\fi
\ifx \citeauthoryear \undefined \def \citeauthoryear#1{#1}\fi
\ifx \endbibitem  \undefined \def \endbibitem {}\fi
\ifx \bconflocation  \undefined \def \bconflocation#1{#1}\fi
\ifx \arxivurl  \undefined \def \arxivurl#1{\textsf{#1}}\fi
\csname PreBibitemsHook\endcsname

\bibitem[\protect\citeauthoryear{Karplus and
  McCammon}{2002}]{karplus_molecular_2002}
\begin{barticle}
\bauthor{\bsnm{Karplus}, \binits{M.}},
\bauthor{\bsnm{McCammon}, \binits{J.A.}}:
\batitle{Molecular dynamics simulations of biomolecules}.
\bjtitle{Nature Structural Biology}
\bvolume{9}(\bissue{9}),
\bfpage{646}--\blpage{652}
(\byear{2002})
\doiurl{10.1038/nsb0902-646}
\end{barticle}
\endbibitem

\bibitem[\protect\citeauthoryear{Dror
  et~al.}{2012}]{BiomolecularSimulation_Dror_2012}
\begin{barticle}
\bauthor{\bsnm{Dror}, \binits{R.O.}},
\bauthor{\bsnm{Dirks}, \binits{R.M.}},
\bauthor{\bsnm{Grossman}, \binits{J.P.}},
\bauthor{\bsnm{Xu}, \binits{H.}},
\bauthor{\bsnm{Shaw}, \binits{D.E.}}:
\batitle{Biomolecular simulation: A computational microscope for molecular
  biology}.
\bjtitle{Annual Review of Biophysics}
\bvolume{41}(\bissue{Volume 41, 2012}),
\bfpage{429}--\blpage{452}
(\byear{2012})
\doiurl{10.1146/annurev-biophys-042910-155245}
\end{barticle}
\endbibitem

\bibitem[\protect\citeauthoryear{Hollingsworth and
  Dror}{2018}]{MolecularDynamics_Scott_2018}
\begin{barticle}
\bauthor{\bsnm{Hollingsworth}, \binits{S.A.}},
\bauthor{\bsnm{Dror}, \binits{R.O.}}:
\batitle{Molecular dynamics simulation for all}.
\bjtitle{Neuron}
\bvolume{99}(\bissue{6}),
\bfpage{1129}--\blpage{1143}
(\byear{2018})
\doiurl{10.1016/j.neuron.2018.08.011}
\end{barticle}
\endbibitem

\bibitem[\protect\citeauthoryear{Durrant and
  McCammon}{2011}]{durrant_molecular_2011}
\begin{barticle}
\bauthor{\bsnm{Durrant}, \binits{J.D.}},
\bauthor{\bsnm{McCammon}, \binits{J.A.}}:
\batitle{Molecular dynamics simulations and drug discovery}.
\bjtitle{BMC Biology}
\bvolume{9}(\bissue{1}),
\bfpage{71}
(\byear{2011})
\doiurl{10.1186/1741-7007-9-71}
\end{barticle}
\endbibitem

\bibitem[\protect\citeauthoryear{De~Vivo
  et~al.}{2016}]{DeVivo_RoleofMolecularDynamics_2016}
\begin{barticle}
\bauthor{\bsnm{De~Vivo}, \binits{M.}},
\bauthor{\bsnm{Masetti}, \binits{M.}},
\bauthor{\bsnm{Bottegoni}, \binits{G.}},
\bauthor{\bsnm{Cavalli}, \binits{A.}}:
\batitle{Role of molecular dynamics and related methods in drug discovery}.
\bjtitle{Journal of Medicinal Chemistry}
\bvolume{59}(\bissue{9}),
\bfpage{4035}--\blpage{4061}
(\byear{2016})
\doiurl{10.1021/acs.jmedchem.5b01684}
{\href{https://arxiv.org/abs/https://doi.org/10.1021/acs.jmedchem.5b01684}{{https://doi.org/10.1021/acs.jmedchem.5b01684}}}.
\bcomment{PMID: 26807648}
\end{barticle}
\endbibitem

\bibitem[\protect\citeauthoryear{Ponder and
  Case}{2003}]{Ponder_ForceFieldsforProteinSimulations_2003}
\begin{bchapter}
\bauthor{\bsnm{Ponder}, \binits{J.W.}},
\bauthor{\bsnm{Case}, \binits{D.A.}}:
\bctitle{Force fields for protein simulations}.
In: \bbtitle{Protein Simulations}.
\bsertitle{Advances in Protein Chemistry},
vol. \bseriesno{66},
pp. \bfpage{27}--\blpage{85}.
\bpublisher{Academic Press}, \blocation{???}
(\byear{2003}).
\doiurl{10.1016/S0065-3233(03)66002-X} .
\burl{https://www.sciencedirect.com/science/article/pii/S006532330366002X}
\end{bchapter}
\endbibitem

\bibitem[\protect\citeauthoryear{Mackerell~Jr.}{2004}]{Mackerell_Empiricalforcefieldsforbiological_2004}
\begin{barticle}
\bauthor{\bsnm{Mackerell~Jr.}, \binits{A.D.}}:
\batitle{Empirical force fields for biological macromolecules: Overview and
  issues}.
\bjtitle{Journal of Computational Chemistry}
\bvolume{25}(\bissue{13}),
\bfpage{1584}--\blpage{1604}
(\byear{2004})
\doiurl{10.1002/jcc.20082}
{\href{https://arxiv.org/abs/https://onlinelibrary.wiley.com/doi/pdf/10.1002/jcc.20082}{{https://onlinelibrary.wiley.com/doi/pdf/10.1002/jcc.20082}}}
\end{barticle}
\endbibitem

\bibitem[\protect\citeauthoryear{Lopes
  et~al.}{2015}]{Lopes_CurrentStatusofProteinForceFields_2025}
\begin{bchapter}
\bauthor{\bsnm{Lopes}, \binits{P.E.M.}},
\bauthor{\bsnm{Guvench}, \binits{O.}},
\bauthor{\bsnm{MacKerell}, \binits{A.D.}}:
\bctitle{Current {Status} of {Protein} {Force} {Fields} for {Molecular}
  {Dynamics} {Simulations}}.
In: \beditor{\bsnm{Kukol}, \binits{A.}} (ed.)
\bbtitle{Molecular {Modeling} of {Proteins}},
pp. \bfpage{47}--\blpage{71}.
\bpublisher{Springer},
\blocation{New York, NY}
(\byear{2015}).
\doiurl{10.1007/978-1-4939-1465-4_3}
\end{bchapter}
\endbibitem

\bibitem[\protect\citeauthoryear{{van der
  Spoel}}{2021}]{Spoel_Systematicdesignofbiomolecularforcefields_2021}
\begin{barticle}
\bauthor{\bsnm{{van der Spoel}}, \binits{D.}}:
\batitle{Systematic design of biomolecular force fields}.
\bjtitle{Current Opinion in Structural Biology}
\bvolume{67},
\bfpage{18}--\blpage{24}
(\byear{2021})
\doiurl{10.1016/j.sbi.2020.08.006} .
\bcomment{Theory and Simulation/Computational Methods Macromolecular
  Assemblies}
\end{barticle}
\endbibitem

\bibitem[\protect\citeauthoryear{Kabylda
  et~al.}{2026}]{kabylda_atomsinteractmolecules_2026}
\begin{botherref}
\oauthor{\bsnm{Kabylda}, \binits{A.}},
\oauthor{\bsnm{Esders}, \binits{M.}},
\oauthor{\bsnm{Gori}, \binits{M.}},
\oauthor{\bsnm{Chmiela}, \binits{S.}},
\oauthor{\bsnm{Müller}, \binits{K.-R.}},
\oauthor{\bsnm{Tkatchenko}, \binits{A.}}:
How Atoms Interact Within Molecules
(2026).
\url{https://arxiv.org/abs/2605.28960}
\end{botherref}
\endbibitem

\bibitem[\protect\citeauthoryear{Störmer and
  Zavadlav}{2026}]{stoermer_aluminumsolidificationnanopolycrystaldeformation_2026}
\begin{botherref}
\oauthor{\bsnm{Störmer}, \binits{I.}},
\oauthor{\bsnm{Zavadlav}, \binits{J.}}:
Aluminum solidification and nanopolycrystal deformation via a Graph Neural
  Network Potential and Million-Atom Simulations
(2026).
\url{https://arxiv.org/abs/2603.24360}
\end{botherref}
\endbibitem

\bibitem[\protect\citeauthoryear{Fuchs
  et~al.}{2025}]{Fuchs_chemtrainDeploy_2025}
\begin{barticle}
\bauthor{\bsnm{Fuchs}, \binits{P.}},
\bauthor{\bsnm{Chen}, \binits{W.}},
\bauthor{\bsnm{Thaler}, \binits{S.}},
\bauthor{\bsnm{Zavadlav}, \binits{J.}}:
\batitle{chemtrain-deploy: A parallel and scalable framework for machine
  learning potentials in million-atom md simulations}.
\bjtitle{Journal of Chemical Theory and Computation}
\bvolume{21}(\bissue{15}),
\bfpage{7550}--\blpage{7560}
(\byear{2025})
\doiurl{10.1021/acs.jctc.5c00996}
{\href{https://arxiv.org/abs/https://doi.org/10.1021/acs.jctc.5c00996}{{https://doi.org/10.1021/acs.jctc.5c00996}}}.
\bcomment{PMID: 40699940}
\end{barticle}
\endbibitem

\bibitem[\protect\citeauthoryear{Batatia et~al.}{2022}]{Batatia_mace_2022}
\begin{bchapter}
\bauthor{\bsnm{Batatia}, \binits{I.}},
\bauthor{\bsnm{Kovacs}, \binits{D.P.}},
\bauthor{\bsnm{Simm}, \binits{G.N.C.}},
\bauthor{\bsnm{Ortner}, \binits{C.}},
\bauthor{\bsnm{Csanyi}, \binits{G.}}:
\bctitle{{MACE}: Higher order equivariant message passing neural networks for
  fast and accurate force fields}.
In: \beditor{\bsnm{Oh}, \binits{A.H.}},
\beditor{\bsnm{Agarwal}, \binits{A.}},
\beditor{\bsnm{Belgrave}, \binits{D.}},
\beditor{\bsnm{Cho}, \binits{K.}} (eds.)
\bbtitle{Advances in Neural Information Processing Systems}
(\byear{2022}).
\burl{https://openreview.net/forum?id=YPpSngE-ZU}
\end{bchapter}
\endbibitem

\bibitem[\protect\citeauthoryear{Levine et~al.}{2026}]{levine_omol_2025}
\begin{botherref}
\oauthor{\bsnm{Levine}, \binits{D.S.}},
\oauthor{\bsnm{Shuaibi}, \binits{M.}},
\oauthor{\bsnm{Spotte-Smith}, \binits{E.W.C.}},
\oauthor{\bsnm{Taylor}, \binits{M.G.}},
\oauthor{\bsnm{Hasyim}, \binits{M.R.}},
\oauthor{\bsnm{Michel}, \binits{K.}},
\oauthor{\bsnm{Batatia}, \binits{I.}},
\oauthor{\bsnm{Csányi}, \binits{G.}},
\oauthor{\bsnm{Dzamba}, \binits{M.}},
\oauthor{\bsnm{Eastman}, \binits{P.}},
\oauthor{\bsnm{Frey}, \binits{N.C.}},
\oauthor{\bsnm{Fu}, \binits{X.}},
\oauthor{\bsnm{Gharakhanyan}, \binits{V.}},
\oauthor{\bsnm{Krishnapriyan}, \binits{A.S.}},
\oauthor{\bsnm{Rackers}, \binits{J.A.}},
\oauthor{\bsnm{Raja}, \binits{S.}},
\oauthor{\bsnm{Rizvi}, \binits{A.}},
\oauthor{\bsnm{Rosen}, \binits{A.S.}},
\oauthor{\bsnm{Ulissi}, \binits{Z.}},
\oauthor{\bsnm{Vargas}, \binits{S.}},
\oauthor{\bsnm{Zitnick}, \binits{C.L.}},
\oauthor{\bsnm{Blau}, \binits{S.M.}},
\oauthor{\bsnm{Wood}, \binits{B.M.}}:
The Open Molecules 2025 (OMol25) Dataset, Evaluations, and Models
(2026).
\url{https://arxiv.org/abs/2505.08762}
\end{botherref}
\endbibitem

\bibitem[\protect\citeauthoryear{Kabylda et~al.}{2026}]{kabylda_qcell_2026}
\begin{botherref}
\oauthor{\bsnm{Kabylda}, \binits{A.}},
\oauthor{\bsnm{Suárez-Dou}, \binits{S.}},
\oauthor{\bsnm{Davoine}, \binits{N.}},
\oauthor{\bsnm{Brünig}, \binits{F.N.}},
\oauthor{\bsnm{Tkatchenko}, \binits{A.}}:
QCell: Comprehensive Quantum-Mechanical Dataset Spanning Diverse Biomolecular
  Fragments
(2026).
\url{https://arxiv.org/abs/2510.09939}
\end{botherref}
\endbibitem

\bibitem[\protect\citeauthoryear{Eastman et~al.}{2023}]{eastman_spice_2023}
\begin{barticle}
\bauthor{\bsnm{Eastman}, \binits{P.}},
\bauthor{\bsnm{Behara}, \binits{P.K.}},
\bauthor{\bsnm{Dotson}, \binits{D.L.}},
\bauthor{\bsnm{Galvelis}, \binits{R.}},
\bauthor{\bsnm{Herr}, \binits{J.E.}},
\bauthor{\bsnm{Horton}, \binits{J.T.}},
\bauthor{\bsnm{Mao}, \binits{Y.}},
\bauthor{\bsnm{Chodera}, \binits{J.D.}},
\bauthor{\bsnm{Pritchard}, \binits{B.P.}},
\bauthor{\bsnm{Wang}, \binits{Y.}},
\bauthor{\bsnm{De~Fabritiis}, \binits{G.}},
\bauthor{\bsnm{Markland}, \binits{T.E.}}:
\batitle{{SPICE}, {A} {Dataset} of {Drug}-like {Molecules} and {Peptides} for
  {Training} {Machine} {Learning} {Potentials}}.
\bjtitle{Scientific Data}
\bvolume{10}(\bissue{1}),
\bfpage{11}
(\byear{2023})
\doiurl{10.1038/s41597-022-01882-6}
\end{barticle}
\endbibitem

\bibitem[\protect\citeauthoryear{Kovács et~al.}{2025}]{kovacs_2025_maceoff}
\begin{barticle}
\bauthor{\bsnm{Kovács}, \binits{D.P.}},
\bauthor{\bsnm{Moore}, \binits{J.H.}},
\bauthor{\bsnm{Browning}, \binits{N.J.}},
\bauthor{\bsnm{Batatia}, \binits{I.}},
\bauthor{\bsnm{Horton}, \binits{J.T.}},
\bauthor{\bsnm{Pu}, \binits{Y.}},
\bauthor{\bsnm{Kapil}, \binits{V.}},
\bauthor{\bsnm{Witt}, \binits{W.C.}},
\bauthor{\bsnm{Magdău}, \binits{I.-B.}},
\bauthor{\bsnm{Cole}, \binits{D.J.}},
\bauthor{\bsnm{Csányi}, \binits{G.}}:
\batitle{Mace-off: Short-range transferable machine learning force fields for
  organic molecules}.
\bjtitle{Journal of the American Chemical Society}
\bvolume{147}(\bissue{21}),
\bfpage{17598}--\blpage{17611}
(\byear{2025})
\doiurl{10.1021/jacs.4c07099}
{\href{https://arxiv.org/abs/https://doi.org/10.1021/jacs.4c07099}{{https://doi.org/10.1021/jacs.4c07099}}}.
\bcomment{PMID: 40387214}
\end{barticle}
\endbibitem

\bibitem[\protect\citeauthoryear{Kabylda et~al.}{2025}]{Kabylda_so3lr_2025}
\begin{barticle}
\bauthor{\bsnm{Kabylda}, \binits{A.}},
\bauthor{\bsnm{Frank}, \binits{J.T.}},
\bauthor{\bsnm{Suárez-Dou}, \binits{S.}},
\bauthor{\bsnm{Khabibrakhmanov}, \binits{A.}},
\bauthor{\bsnm{Medrano~Sandonas}, \binits{L.}},
\bauthor{\bsnm{Unke}, \binits{O.T.}},
\bauthor{\bsnm{Chmiela}, \binits{S.}},
\bauthor{\bsnm{M{\"u}ller}, \binits{K.-R.}},
\bauthor{\bsnm{Tkatchenko}, \binits{A.}}:
\batitle{Molecular simulations with a pretrained neural network and universal
  pairwise force fields}.
\bjtitle{Journal of the American Chemical Society}
\bvolume{147}(\bissue{37}),
\bfpage{33723}--\blpage{33734}
(\byear{2025})
\doiurl{10.1021/jacs.5c09558}
{\href{https://arxiv.org/abs/https://doi.org/10.1021/jacs.5c09558}{{https://doi.org/10.1021/jacs.5c09558}}}.
\bcomment{PMID: 40886167}
\end{barticle}
\endbibitem

\bibitem[\protect\citeauthoryear{Unke et~al.}{2024}]{Unke_gems_2024}
\begin{barticle}
\bauthor{\bsnm{Unke}, \binits{O.T.}},
\bauthor{\bsnm{Stöhr}, \binits{M.}},
\bauthor{\bsnm{Ganscha}, \binits{S.}},
\bauthor{\bsnm{Unterthiner}, \binits{T.}},
\bauthor{\bsnm{Maennel}, \binits{H.}},
\bauthor{\bsnm{Kashubin}, \binits{S.}},
\bauthor{\bsnm{Ahlin}, \binits{D.}},
\bauthor{\bsnm{Gastegger}, \binits{M.}},
\bauthor{\bsnm{Sandonas}, \binits{L.M.}},
\bauthor{\bsnm{Berryman}, \binits{J.T.}},
\bauthor{\bsnm{Tkatchenko}, \binits{A.}},
\bauthor{\bsnm{Müller}, \binits{K.-R.}}:
\batitle{Biomolecular dynamics with machine-learned quantum-mechanical force
  fields trained on diverse chemical fragments}.
\bjtitle{Science Advances}
\bvolume{10}(\bissue{14}),
\bfpage{4397}
(\byear{2024})
\doiurl{10.1126/sciadv.adn4397}
{\href{https://arxiv.org/abs/https://www.science.org/doi/pdf/10.1126/sciadv.adn4397}{{https://www.science.org/doi/pdf/10.1126/sciadv.adn4397}}}
\end{barticle}
\endbibitem

\bibitem[\protect\citeauthoryear{Hénin et~al.}{2022}]{henin_enhanced_2022}
\begin{barticle}
\bauthor{\bsnm{Hénin}, \binits{J.}},
\bauthor{\bsnm{Lelièvre}, \binits{T.}},
\bauthor{\bsnm{Shirts}, \binits{M.R.}},
\bauthor{\bsnm{Valsson}, \binits{O.}},
\bauthor{\bsnm{Delemotte}, \binits{L.}}:
\batitle{Enhanced {Sampling} {Methods} for {Molecular} {Dynamics} {Simulations}
  [{Article} v1.0]}.
\bjtitle{Living Journal of Computational Molecular Science}
\bvolume{4}(\bissue{1}),
\bfpage{1583}
(\byear{2022})
\doiurl{10.33011/livecoms.4.1.1583} .
\bcomment{Chap. Articles}.
Accessed 2026-05-07
\end{barticle}
\endbibitem

\bibitem[\protect\citeauthoryear{Mehdi
  et~al.}{2024}]{Mehdi_EnhancedSampling_2024}
\begin{barticle}
\bauthor{\bsnm{Mehdi}, \binits{S.}},
\bauthor{\bsnm{Smith}, \binits{Z.}},
\bauthor{\bsnm{Herron}, \binits{L.}},
\bauthor{\bsnm{Zou}, \binits{Z.}},
\bauthor{\bsnm{Tiwary}, \binits{P.}}:
\batitle{Enhanced sampling with machine learning}.
\bjtitle{Annual Review of Physical Chemistry}
\bvolume{75}(\bissue{Volume 75, 2024}),
\bfpage{347}--\blpage{370}
(\byear{2024})
\doiurl{10.1146/annurev-physchem-083122-125941}
\end{barticle}
\endbibitem

\bibitem[\protect\citeauthoryear{Zhu et~al.}{2026}]{Zhu_EnhancedSampling_2026}
\begin{barticle}
\bauthor{\bsnm{Zhu}, \binits{K.}},
\bauthor{\bsnm{Trizio}, \binits{E.}},
\bauthor{\bsnm{Zhang}, \binits{J.}},
\bauthor{\bsnm{Hu}, \binits{R.}},
\bauthor{\bsnm{Jiang}, \binits{L.}},
\bauthor{\bsnm{Hou}, \binits{T.}},
\bauthor{\bsnm{Bonati}, \binits{L.}}:
\batitle{Enhanced sampling in the age of machine learning: Algorithms and
  applications}.
\bjtitle{Chemical Reviews}
\bvolume{126}(\bissue{1}),
\bfpage{671}--\blpage{713}
(\byear{2026})
\doiurl{10.1021/acs.chemrev.5c00700}
{\href{https://arxiv.org/abs/https://doi.org/10.1021/acs.chemrev.5c00700}{{https://doi.org/10.1021/acs.chemrev.5c00700}}}.
\bcomment{PMID: 41124671}
\end{barticle}
\endbibitem

\bibitem[\protect\citeauthoryear{Noid}{2013}]{Noid_gbbio_2013}
\begin{barticle}
\bauthor{\bsnm{Noid}, \binits{W.G.}}:
\batitle{Perspective: {Coarse}-grained models for biomolecular systems}.
\bjtitle{The Journal of Chemical Physics}
\bvolume{139}(\bissue{9}),
\bfpage{090901}
(\byear{2013})
\doiurl{10.1063/1.4818908} .
Accessed 2026-07-05
\end{barticle}
\endbibitem

\bibitem[\protect\citeauthoryear{Singh and Li}{2019}]{Singh_cgmodels_2019}
\begin{barticle}
\bauthor{\bsnm{Singh}, \binits{N.}},
\bauthor{\bsnm{Li}, \binits{W.}}:
\batitle{Recent {Advances} in {Coarse}-{Grained} {Models} for {Biomolecules}
  and {Their} {Applications}}.
\bjtitle{International Journal of Molecular Sciences}
\bvolume{20}(\bissue{15}),
\bfpage{3774}
(\byear{2019})
\doiurl{10.3390/ijms20153774}
\end{barticle}
\endbibitem

\bibitem[\protect\citeauthoryear{Airas et~al.}{2023}]{Airas_transferableIsGNN}
\begin{barticle}
\bauthor{\bsnm{Airas}, \binits{J.}},
\bauthor{\bsnm{Ding}, \binits{X.}},
\bauthor{\bsnm{Zhang}, \binits{B.}}:
\batitle{Transferable implicit solvation via contrastive learning of graph
  neural networks}.
\bjtitle{ACS Central Science}
\bvolume{9}(\bissue{12}),
\bfpage{2286}--\blpage{2297}
(\byear{2023})
\doiurl{10.1021/acscentsci.3c01160}
{\href{https://arxiv.org/abs/https://doi.org/10.1021/acscentsci.3c01160}{{https://doi.org/10.1021/acscentsci.3c01160}}}
\end{barticle}
\endbibitem

\bibitem[\protect\citeauthoryear{Ding and Zhang}{2022}]{Ding_clcgff_2022}
\begin{barticle}
\bauthor{\bsnm{Ding}, \binits{X.}},
\bauthor{\bsnm{Zhang}, \binits{B.}}:
\batitle{Contrastive learning of coarse-grained force fields}.
\bjtitle{Journal of Chemical Theory and Computation}
\bvolume{18}(\bissue{10}),
\bfpage{6334}--\blpage{6344}
(\byear{2022})
\doiurl{10.1021/acs.jctc.2c00616}
{\href{https://arxiv.org/abs/https://doi.org/10.1021/acs.jctc.2c00616}{{https://doi.org/10.1021/acs.jctc.2c00616}}}.
\bcomment{PMID: 36112935}
\end{barticle}
\endbibitem

\bibitem[\protect\citeauthoryear{Onufriev and
  Case}{2019}]{Onufriev_gnnbis_2019}
\begin{barticle}
\bauthor{\bsnm{Onufriev}, \binits{A.V.}},
\bauthor{\bsnm{Case}, \binits{D.A.}}:
\batitle{Generalized born implicit solvent models for biomolecules}.
\bjtitle{Annual Review of Biophysics}
\bvolume{48}(\bissue{Volume 48, 2019}),
\bfpage{275}--\blpage{296}
(\byear{2019})
\doiurl{10.1146/annurev-biophys-052118-115325}
\end{barticle}
\endbibitem

\bibitem[\protect\citeauthoryear{Chen et~al.}{2021}]{chen_machine_2021}
\begin{barticle}
\bauthor{\bsnm{Chen}, \binits{Y.}},
\bauthor{\bsnm{Krämer}, \binits{A.}},
\bauthor{\bsnm{Charron}, \binits{N.E.}},
\bauthor{\bsnm{Husic}, \binits{B.E.}},
\bauthor{\bsnm{Clementi}, \binits{C.}},
\bauthor{\bsnm{Noé}, \binits{F.}}:
\batitle{Machine learning implicit solvation for molecular dynamics}.
\bjtitle{The Journal of Chemical Physics}
\bvolume{155}(\bissue{8}),
\bfpage{084101}
(\byear{2021})
\doiurl{10.1063/5.0059915} .
Accessed 2026-07-05
\end{barticle}
\endbibitem

\bibitem[\protect\citeauthoryear{Majewski et~al.}{2023}]{majewski_machine_2023}
\begin{barticle}
\bauthor{\bsnm{Majewski}, \binits{M.}},
\bauthor{\bsnm{Pérez}, \binits{A.}},
\bauthor{\bsnm{Thölke}, \binits{P.}},
\bauthor{\bsnm{Doerr}, \binits{S.}},
\bauthor{\bsnm{Charron}, \binits{N.E.}},
\bauthor{\bsnm{Giorgino}, \binits{T.}},
\bauthor{\bsnm{Husic}, \binits{B.E.}},
\bauthor{\bsnm{Clementi}, \binits{C.}},
\bauthor{\bsnm{Noé}, \binits{F.}},
\bauthor{\bsnm{De~Fabritiis}, \binits{G.}}:
\batitle{Machine learning coarse-grained potentials of protein thermodynamics}.
\bjtitle{Nature Communications}
\bvolume{14}(\bissue{1}),
\bfpage{5739}
(\byear{2023})
\doiurl{10.1038/s41467-023-41343-1}
\end{barticle}
\endbibitem

\bibitem[\protect\citeauthoryear{Wang et~al.}{2019}]{Wang_mlcgff_2019}
\begin{barticle}
\bauthor{\bsnm{Wang}, \binits{J.}},
\bauthor{\bsnm{Olsson}, \binits{S.}},
\bauthor{\bsnm{Wehmeyer}, \binits{C.}},
\bauthor{\bsnm{P{\'e}rez}, \binits{A.}},
\bauthor{\bsnm{Charron}, \binits{N.E.}},
\bauthor{\bsnm{Fabritiis}, \binits{G.}},
\bauthor{\bsnm{No{\'e}}, \binits{F.}},
\bauthor{\bsnm{Clementi}, \binits{C.}}:
\batitle{Machine learning of coarse-grained molecular dynamics force fields}.
\bjtitle{ACS Central Science}
\bvolume{5}(\bissue{5}),
\bfpage{755}--\blpage{767}
(\byear{2019})
\doiurl{10.1021/acscentsci.8b00913}
{\href{https://arxiv.org/abs/https://doi.org/10.1021/acscentsci.8b00913}{{https://doi.org/10.1021/acscentsci.8b00913}}}.
\bcomment{PMID: 31139712}
\end{barticle}
\endbibitem

\bibitem[\protect\citeauthoryear{Noid
  et~al.}{2024}]{Noid_RigorousProgressinCoarseGraining_2024}
\begin{barticle}
\bauthor{\bsnm{Noid}, \binits{W.G.}},
\bauthor{\bsnm{Szukalo}, \binits{R.J.}},
\bauthor{\bsnm{Kidder}, \binits{K.M.}},
\bauthor{\bsnm{Lesniewski}, \binits{M.C.}}:
\batitle{Rigorous progress in coarse-graining}.
\bjtitle{Annual Review of Physical Chemistry}
\bvolume{75}(\bissue{Volume 75, 2024}),
\bfpage{21}--\blpage{45}
(\byear{2024})
\doiurl{10.1146/annurev-physchem-062123-010821}
\end{barticle}
\endbibitem

\bibitem[\protect\citeauthoryear{Slejko et~al.}{2026}]{Slejko2026}
\begin{botherref}
\oauthor{\bsnm{Slejko}, \binits{E.}},
\oauthor{\bsnm{Coste}, \binits{A.}},
\oauthor{\bsnm{Potisk}, \binits{T.}},
\oauthor{\bsnm{Zavadlav}, \binits{J.}},
\oauthor{\bsnm{Praprotnik}, \binits{M.}}:
Achieving all-atom molecular dynamics accuracy from the poisson–boltzmann
  method through machine learning.
The Journal of Chemical Physics
\textbf{164}(5)
(2026)
\doiurl{10.1063/5.0313624}
\end{botherref}
\endbibitem

\bibitem[\protect\citeauthoryear{Coste et~al.}{2023}]{Coste2023}
\begin{barticle}
\bauthor{\bsnm{Coste}, \binits{A.}},
\bauthor{\bsnm{Slejko}, \binits{E.}},
\bauthor{\bsnm{Zavadlav}, \binits{J.}},
\bauthor{\bsnm{Praprotnik}, \binits{M.}}:
\batitle{Developing an implicit solvation machine learning model for molecular
  simulations of ionic media}.
\bjtitle{Journal of Chemical Theory and Computation}
(\byear{2023})
\doiurl{10.1021/acs.jctc.3c00984}
\end{barticle}
\endbibitem

\bibitem[\protect\citeauthoryear{Thaler et~al.}{2022}]{Thaler_REmin_2022}
\begin{barticle}
\bauthor{\bsnm{Thaler}, \binits{S.}},
\bauthor{\bsnm{Stupp}, \binits{M.}},
\bauthor{\bsnm{Zavadlav}, \binits{J.}}:
\batitle{Deep coarse-grained potentials via relative entropy minimization}.
\bjtitle{The Journal of Chemical Physics}
\bvolume{157}(\bissue{24}),
\bfpage{244103}
(\byear{2022})
\doiurl{10.1063/5.0124538} .
Accessed 2026-07-04
\end{barticle}
\endbibitem

\bibitem[\protect\citeauthoryear{Katzberger and
  Riniker}{2024}]{katzberger_2024_ageneralgraph}
\begin{barticle}
\bauthor{\bsnm{Katzberger}, \binits{P.}},
\bauthor{\bsnm{Riniker}, \binits{S.}}:
\batitle{A general graph neural network based implicit solvation model for
  organic molecules in water}.
\bjtitle{Chem. Sci.}
\bvolume{15},
\bfpage{10794}--\blpage{10802}
(\byear{2024})
\doiurl{10.1039/D4SC02432J}
\end{barticle}
\endbibitem

\bibitem[\protect\citeauthoryear{Charron
  et~al.}{2025}]{charron_2023_navigating}
\begin{barticle}
\bauthor{\bsnm{Charron}, \binits{N.E.}},
\bauthor{\bsnm{Bonneau}, \binits{K.}},
\bauthor{\bsnm{{Pasos-Trejo}}, \binits{A.S.}},
\bauthor{\bsnm{Guljas}, \binits{A.}},
\bauthor{\bsnm{Chen}, \binits{Y.}},
\bauthor{\bsnm{Musil}, \binits{F.}},
\bauthor{\bsnm{Venturin}, \binits{J.}},
\bauthor{\bsnm{Gusew}, \binits{D.}},
\bauthor{\bsnm{Zaporozhets}, \binits{I.}},
\bauthor{\bsnm{Kr{\"a}mer}, \binits{A.}},
\bauthor{\bsnm{Templeton}, \binits{C.}},
\bauthor{\bsnm{Kelkar}, \binits{A.}},
\bauthor{\bsnm{Durumeric}, \binits{A.E.P.}},
\bauthor{\bsnm{Olsson}, \binits{S.}},
\bauthor{\bsnm{P{\'e}rez}, \binits{A.}},
\bauthor{\bsnm{Majewski}, \binits{M.}},
\bauthor{\bsnm{Husic}, \binits{B.E.}},
\bauthor{\bsnm{Patel}, \binits{A.}},
\bauthor{\bsnm{De~Fabritiis}, \binits{G.}},
\bauthor{\bsnm{No{\'e}}, \binits{F.}},
\bauthor{\bsnm{Clementi}, \binits{C.}}:
\batitle{Navigating protein landscapes with a machine-learned transferable
  coarse-grained model}.
\bjtitle{Nature Chemistry}
(\byear{2025})
\doiurl{10.1038/s41557-025-01874-0}
\end{barticle}
\endbibitem

\bibitem[\protect\citeauthoryear{Durumeric
  et~al.}{2026}]{durumeric_learning_2026}
\begin{barticle}
\bauthor{\bsnm{Durumeric}, \binits{A.E.P.}},
\bauthor{\bsnm{Chen}, \binits{Y.}},
\bauthor{\bsnm{Pasos-Trejo}, \binits{A.S.}},
\bauthor{\bsnm{Noé}, \binits{F.}},
\bauthor{\bsnm{Clementi}, \binits{C.}}:
\batitle{Learning data-efficient coarse-grained molecular dynamics from forces
  and noise}.
\bjtitle{Nature Communications}
\bvolume{17}(\bissue{1}),
\bfpage{2493}
(\byear{2026})
\doiurl{10.1038/s41467-026-70818-0}
\end{barticle}
\endbibitem

\bibitem[\protect\citeauthoryear{Chen
  et~al.}{2026}]{Chen_EnhancedSampling_2025}
\begin{barticle}
\bauthor{\bsnm{Chen}, \binits{W.}},
\bauthor{\bsnm{G{\"o}rlich}, \binits{F.}},
\bauthor{\bsnm{Fuchs}, \binits{P.}},
\bauthor{\bsnm{Zavadlav}, \binits{J.}}:
\batitle{Enhanced sampling for efficient learning of coarse-grained machine
  learning potentials}.
\bjtitle{Journal of Chemical Theory and Computation}
\bvolume{22}(\bissue{1}),
\bfpage{219}--\blpage{230}
(\byear{2026})
\doiurl{10.1021/acs.jctc.5c01712}
{\href{https://arxiv.org/abs/https://doi.org/10.1021/acs.jctc.5c01712}{{https://doi.org/10.1021/acs.jctc.5c01712}}}.
\bcomment{PMID: 41437682}
\end{barticle}
\endbibitem

\bibitem[\protect\citeauthoryear{Ding and Zhang}{2022}]{ding2022contrastive}
\begin{barticle}
\bauthor{\bsnm{Ding}, \binits{X.}},
\bauthor{\bsnm{Zhang}, \binits{B.}}:
\batitle{Contrastive learning of coarse-grained force fields}.
\bjtitle{Journal of chemical theory and computation}
\bvolume{18}(\bissue{10}),
\bfpage{6334}--\blpage{6344}
(\byear{2022})
\end{barticle}
\endbibitem

\bibitem[\protect\citeauthoryear{Airas et~al.}{2023}]{airas2023transferable}
\begin{barticle}
\bauthor{\bsnm{Airas}, \binits{J.}},
\bauthor{\bsnm{Ding}, \binits{X.}},
\bauthor{\bsnm{Zhang}, \binits{B.}}:
\batitle{Transferable implicit solvation via contrastive learning of graph
  neural networks}.
\bjtitle{ACS Central Science}
\bvolume{9}(\bissue{12}),
\bfpage{2286}--\blpage{2297}
(\byear{2023})
\end{barticle}
\endbibitem

\bibitem[\protect\citeauthoryear{Klamt}{1995}]{klamt1995conductor}
\begin{barticle}
\bauthor{\bsnm{Klamt}, \binits{A.}}:
\batitle{Conductor-like screening model for real solvents: a new approach to
  the quantitative calculation of solvation phenomena}.
\bjtitle{The Journal of Physical Chemistry}
\bvolume{99}(\bissue{7}),
\bfpage{2224}--\blpage{2235}
(\byear{1995})
\end{barticle}
\endbibitem

\bibitem[\protect\citeauthoryear{Barone et~al.}{2013}]{Barone_pcm_2013}
\begin{barticle}
\bauthor{\bsnm{Barone}, \binits{V.}},
\bauthor{\bsnm{Carnimeo}, \binits{I.}},
\bauthor{\bsnm{Scalmani}, \binits{G.}}:
\batitle{Computational spectroscopy of large systems in solution: The dftb/pcm
  and td-dftb/pcm approach}.
\bjtitle{Journal of Chemical Theory and Computation}
\bvolume{9}(\bissue{4}),
\bfpage{2052}--\blpage{2071}
(\byear{2013})
\doiurl{10.1021/ct301050x}
{\href{https://arxiv.org/abs/https://doi.org/10.1021/ct301050x}{{https://doi.org/10.1021/ct301050x}}}.
\bcomment{PMID: 26583552}
\end{barticle}
\endbibitem

\bibitem[\protect\citeauthoryear{Hou et~al.}{2010}]{Hou_smd_2010}
\begin{barticle}
\bauthor{\bsnm{Hou}, \binits{G.}},
\bauthor{\bsnm{Zhu}, \binits{X.}},
\bauthor{\bsnm{Cui}, \binits{Q.}}:
\batitle{An implicit solvent model for scc-dftb with charge-dependent radii}.
\bjtitle{Journal of Chemical Theory and Computation}
\bvolume{6}(\bissue{8}),
\bfpage{2303}--\blpage{2314}
(\byear{2010})
\doiurl{10.1021/ct1001818}
{\href{https://arxiv.org/abs/https://doi.org/10.1021/ct1001818}{{https://doi.org/10.1021/ct1001818}}}.
\bcomment{PMID: 20711513}
\end{barticle}
\endbibitem

\bibitem[\protect\citeauthoryear{R{\"o}cken
  et~al.}{2024}]{roecken_2024_predictingsolvationFE}
\begin{barticle}
\bauthor{\bsnm{R{\"o}cken}, \binits{S.}},
\bauthor{\bsnm{Burnet}, \binits{A.F.}},
\bauthor{\bsnm{Zavadlav}, \binits{J.}}:
\batitle{Predicting solvation free energies with an implicit solvent machine
  learning potential}.
\bjtitle{The Journal of Chemical Physics}
\bvolume{161}(\bissue{23}),
\bfpage{234101}
(\byear{2024})
\doiurl{10.1063/5.0235189}
\end{barticle}
\endbibitem

\bibitem[\protect\citeauthoryear{Batatia et~al.}{2022}]{batatia_2022_mace}
\begin{bchapter}
\bauthor{\bsnm{Batatia}, \binits{I.}},
\bauthor{\bsnm{Kovacs}, \binits{D.P.}},
\bauthor{\bsnm{Simm}, \binits{G.N.C.}},
\bauthor{\bsnm{Ortner}, \binits{C.}},
\bauthor{\bsnm{Csanyi}, \binits{G.}}:
\bctitle{{MACE}: Higher order equivariant message passing neural networks for
  fast and accurate force fields}.
In: \beditor{\bsnm{Oh}, \binits{A.H.}},
\beditor{\bsnm{Agarwal}, \binits{A.}},
\beditor{\bsnm{Belgrave}, \binits{D.}},
\beditor{\bsnm{Cho}, \binits{K.}} (eds.)
\bbtitle{Advances in Neural Information Processing Systems}
(\byear{2022}).
\burl{https://openreview.net/forum?id=YPpSngE-ZU}
\end{bchapter}
\endbibitem

\bibitem[\protect\citeauthoryear{Eastman et~al.}{2024}]{spice2}
\begin{barticle}
\bauthor{\bsnm{Eastman}, \binits{P.}},
\bauthor{\bsnm{Pritchard}, \binits{B.P.}},
\bauthor{\bsnm{Chodera}, \binits{J.D.}},
\bauthor{\bsnm{Markland}, \binits{T.E.}}:
\batitle{Nutmeg and spice: Models and data for biomolecular machine learning}.
\bjtitle{Journal of Chemical Theory and Computation}
\bvolume{20}(\bissue{19}),
\bfpage{8583}--\blpage{8593}
(\byear{2024})
\doiurl{10.1021/acs.jctc.4c00794}
{\href{https://arxiv.org/abs/https://doi.org/10.1021/acs.jctc.4c00794}{{https://doi.org/10.1021/acs.jctc.4c00794}}}.
\bcomment{PMID: 39318326}
\end{barticle}
\endbibitem

\bibitem[\protect\citeauthoryear{Sillitoe et~al.}{2015}]{Sillitoe_cath_2015}
\begin{barticle}
\bauthor{\bsnm{Sillitoe}, \binits{I.}},
\bauthor{\bsnm{Lewis}, \binits{T.E.}},
\bauthor{\bsnm{Cuff}, \binits{A.}},
\bauthor{\bsnm{Das}, \binits{S.}},
\bauthor{\bsnm{Ashford}, \binits{P.}},
\bauthor{\bsnm{Dawson}, \binits{N.L.}},
\bauthor{\bsnm{Furnham}, \binits{N.}},
\bauthor{\bsnm{Laskowski}, \binits{R.A.}},
\bauthor{\bsnm{Lee}, \binits{D.}},
\bauthor{\bsnm{Lees}, \binits{J.G.}},
\bauthor{\bsnm{Lehtinen}, \binits{S.}},
\bauthor{\bsnm{Studer}, \binits{R.A.}},
\bauthor{\bsnm{Thornton}, \binits{J.}},
\bauthor{\bsnm{Orengo}, \binits{C.A.}}:
\batitle{Cath: comprehensive structural and functional annotations for genome
  sequences}.
\bjtitle{Nucleic Acids Research}
\bvolume{43}(\bissue{D1}),
\bfpage{376}--\blpage{381}
(\byear{2015})
\doiurl{10.1093/nar/gku947}
{\href{https://arxiv.org/abs/https://academic.oup.com/nar/article-pdf/43/D1/D376/7330586/gku947.pdf}{{https://academic.oup.com/nar/article-pdf/43/D1/D376/7330586/gku947.pdf}}}
\end{barticle}
\endbibitem

\bibitem[\protect\citeauthoryear{Mobley and
  Guthrie}{2014}]{mobley_2014_freesolv}
\begin{barticle}
\bauthor{\bsnm{Mobley}, \binits{D.L.}},
\bauthor{\bsnm{Guthrie}, \binits{J.P.}}:
\batitle{Freesolv: a database of experimental and calculated hydration free
  energies, with input files}.
\bjtitle{J. Comput. Aided Mol. Des.}
\bvolume{28}(\bissue{7}),
\bfpage{711}--\blpage{720}
(\byear{2014})
\end{barticle}
\endbibitem

\bibitem[\protect\citeauthoryear{Karwounopoulos
  et~al.}{2024}]{Karwounopoulos_freeEnergyMLP_2024}
\begin{barticle}
\bauthor{\bsnm{Karwounopoulos}, \binits{J.}},
\bauthor{\bsnm{Wu}, \binits{Z.}},
\bauthor{\bsnm{Tkaczyk}, \binits{S.}},
\bauthor{\bsnm{Wang}, \binits{S.}},
\bauthor{\bsnm{Baskerville}, \binits{A.}},
\bauthor{\bsnm{Ranasinghe}, \binits{K.}},
\bauthor{\bsnm{Langer}, \binits{T.}},
\bauthor{\bsnm{Wood}, \binits{G.P.F.}},
\bauthor{\bsnm{Wieder}, \binits{M.}},
\bauthor{\bsnm{Boresch}, \binits{S.}}:
\batitle{Insights and challenges in correcting force field based solvation free
  energies using a neural network potential}.
\bjtitle{The Journal of Physical Chemistry B}
\bvolume{128}(\bissue{28}),
\bfpage{6693}--\blpage{6703}
(\byear{2024})
\doiurl{10.1021/acs.jpcb.4c01417}
{\href{https://arxiv.org/abs/https://doi.org/10.1021/acs.jpcb.4c01417}{{https://doi.org/10.1021/acs.jpcb.4c01417}}}.
\bcomment{PMID: 38976601}
\end{barticle}
\endbibitem

\bibitem[\protect\citeauthoryear{Harry~Moore
  et~al.}{2026}]{Moore_maceoffsolvation_2026}
\begin{barticle}
\bauthor{\bsnm{Harry~Moore}, \binits{J.}},
\bauthor{\bsnm{Cole}, \binits{D.J.}},
\bauthor{\bsnm{Csányi}, \binits{G.}}:
\batitle{Computing solvation free energies of small molecules with experimental
  accuracy}.
\bjtitle{Journal of the American Chemical Society}
\bvolume{148}(\bissue{5}),
\bfpage{4928}--\blpage{4937}
(\byear{2026})
\doiurl{10.1021/jacs.5c10940}
{\href{https://arxiv.org/abs/https://doi.org/10.1021/jacs.5c10940}{{https://doi.org/10.1021/jacs.5c10940}}}.
\bcomment{PMID: 41591329}
\end{barticle}
\endbibitem

\bibitem[\protect\citeauthoryear{Eastman et~al.}{2023}]{spice}
\begin{barticle}
\bauthor{\bsnm{Eastman}, \binits{P.}},
\bauthor{\bsnm{Behara}, \binits{P.K.}},
\bauthor{\bsnm{Dotson}, \binits{D.L.}},
\bauthor{\bsnm{Galvelis}, \binits{R.}},
\bauthor{\bsnm{Herr}, \binits{J.E.}},
\bauthor{\bsnm{Horton}, \binits{J.T.}},
\bauthor{\bsnm{Mao}, \binits{Y.}},
\bauthor{\bsnm{Chodera}, \binits{J.D.}},
\bauthor{\bsnm{Pritchard}, \binits{B.P.}},
\bauthor{\bsnm{Wang}, \binits{Y.}},
\bauthor{\bsnm{De~Fabritiis}, \binits{G.}},
\bauthor{\bsnm{Markland}, \binits{T.E.}}:
\batitle{{SPICE}, {A} {Dataset} of {Drug}-like {Molecules} and {Peptides} for
  {Training} {Machine} {Learning} {Potentials}}.
\bjtitle{Scientific Data}
\bvolume{10}(\bissue{1}),
\bfpage{11}
(\byear{2023})
\doiurl{10.1038/s41597-022-01882-6} .
\bcomment{Publisher: Nature Publishing Group}.
Accessed 2024-10-09
\end{barticle}
\endbibitem

\bibitem[\protect\citeauthoryear{Devereux et~al.}{2020}]{Devereux_ani2x_2020}
\begin{barticle}
\bauthor{\bsnm{Devereux}, \binits{C.}},
\bauthor{\bsnm{Smith}, \binits{J.S.}},
\bauthor{\bsnm{Huddleston}, \binits{K.K.}},
\bauthor{\bsnm{Barros}, \binits{K.}},
\bauthor{\bsnm{Zubatyuk}, \binits{R.}},
\bauthor{\bsnm{Isayev}, \binits{O.}},
\bauthor{\bsnm{Roitberg}, \binits{A.E.}}:
\batitle{Extending the applicability of the ani deep learning molecular
  potential to sulfur and halogens}.
\bjtitle{Journal of Chemical Theory and Computation}
\bvolume{16}(\bissue{7}),
\bfpage{4192}--\blpage{4202}
(\byear{2020})
\doiurl{10.1021/acs.jctc.0c00121}
{\href{https://arxiv.org/abs/https://doi.org/10.1021/acs.jctc.0c00121}{{https://doi.org/10.1021/acs.jctc.0c00121}}}.
\bcomment{PMID: 32543858}
\end{barticle}
\endbibitem

\bibitem[\protect\citeauthoryear{Medders
  et~al.}{2014}]{Medders_FirstPrinciplesWaterPotential_2014}
\begin{barticle}
\bauthor{\bsnm{Medders}, \binits{G.R.}},
\bauthor{\bsnm{Babin}, \binits{V.}},
\bauthor{\bsnm{Paesani}, \binits{F.}}:
\batitle{Development of a “first-principles” water potential with flexible
  monomers. iii. liquid phase properties}.
\bjtitle{Journal of Chemical Theory and Computation}
\bvolume{10}(\bissue{8}),
\bfpage{2906}--\blpage{2910}
(\byear{2014})
\doiurl{10.1021/ct5004115}
{\href{https://arxiv.org/abs/https://doi.org/10.1021/ct5004115}{{https://doi.org/10.1021/ct5004115}}}.
\bcomment{PMID: 26588266}
\end{barticle}
\endbibitem

\bibitem[\protect\citeauthoryear{Soper}{2000}]{Soper_RDFwater_2000}
\begin{barticle}
\bauthor{\bsnm{Soper}, \binits{A.K.}}:
\batitle{The radial distribution functions of water and ice from 220 to 673 k
  and at pressures up to 400 mpa}.
\bjtitle{Chemical Physics}
\bvolume{258}(\bissue{2}),
\bfpage{121}--\blpage{137}
(\byear{2000})
\doiurl{10.1016/S0301-0104(00)00179-8}
\end{barticle}
\endbibitem

\bibitem[\protect\citeauthoryear{Sillitoe et~al.}{2014}]{sillitoe_2014_cath}
\begin{barticle}
\bauthor{\bsnm{Sillitoe}, \binits{I.}},
\bauthor{\bsnm{Lewis}, \binits{T.E.}},
\bauthor{\bsnm{Cuff}, \binits{A.}},
\bauthor{\bsnm{Das}, \binits{S.}},
\bauthor{\bsnm{Ashford}, \binits{P.}},
\bauthor{\bsnm{Dawson}, \binits{N.L.}},
\bauthor{\bsnm{Furnham}, \binits{N.}},
\bauthor{\bsnm{Laskowski}, \binits{R.A.}},
\bauthor{\bsnm{Lee}, \binits{D.}},
\bauthor{\bsnm{Lees}, \binits{J.G.}},
\bauthor{\bsnm{Lehtinen}, \binits{S.}},
\bauthor{\bsnm{Studer}, \binits{R.A.}},
\bauthor{\bsnm{Thornton}, \binits{J.}},
\bauthor{\bsnm{Orengo}, \binits{C.A.}}:
\batitle{Cath: comprehensive structural and functional annotations for genome
  sequences}.
\bjtitle{Nucleic Acids Research}
\bvolume{43}(\bissue{D1}),
\bfpage{376}--\blpage{381}
(\byear{2014})
\doiurl{10.1093/nar/gku947}
{\href{https://arxiv.org/abs/https://academic.oup.com/nar/article-pdf/43/D1/D376/7330586/gku947.pdf}{{https://academic.oup.com/nar/article-pdf/43/D1/D376/7330586/gku947.pdf}}}
\end{barticle}
\endbibitem

\bibitem[\protect\citeauthoryear{Mirarchi
  et~al.}{2024}]{mirarchi_mdcathlargescalemddataset_2024}
\begin{botherref}
\oauthor{\bsnm{Mirarchi}, \binits{A.}},
\oauthor{\bsnm{Giorgino}, \binits{T.}},
\oauthor{\bsnm{Fabritiis}, \binits{G.D.}}:
mdCATH: A Large-Scale MD Dataset for Data-Driven Computational Biophysics
(2024).
\url{https://arxiv.org/abs/2407.14794}
\end{botherref}
\endbibitem

\bibitem[\protect\citeauthoryear{Vermeire and
  Green}{2021}]{Vermeire_combisolv_2021}
\begin{barticle}
\bauthor{\bsnm{Vermeire}, \binits{F.H.}},
\bauthor{\bsnm{Green}, \binits{W.H.}}:
\batitle{Transfer learning for solvation free energies: From quantum chemistry
  to experiments}.
\bjtitle{Chemical Engineering Journal}
\bvolume{418},
\bfpage{129307}
(\byear{2021})
\doiurl{10.1016/j.cej.2021.129307}
\end{barticle}
\endbibitem

\bibitem[\protect\citeauthoryear{Thaler and
  Zavadlav}{2021}]{thaler_2021_learning}
\begin{barticle}
\bauthor{\bsnm{Thaler}, \binits{S.}},
\bauthor{\bsnm{Zavadlav}, \binits{J.}}:
\batitle{Learning neural network potentials from experimental data via
  {Differentiable} {Trajectory} {Reweighting}}.
\bjtitle{Nature Communications}
\bvolume{12}(\bissue{1}),
\bfpage{6884}
(\byear{2021})
\doiurl{10.1038/s41467-021-27241-4} .
\bcomment{Publisher: Nature Publishing Group}.
Accessed 2024-03-05
\end{barticle}
\endbibitem

\bibitem[\protect\citeauthoryear{Zwanzig}{1954}]{Zwanzig_fep_1954}
\begin{barticle}
\bauthor{\bsnm{Zwanzig}, \binits{R.W.}}:
\batitle{High‐temperature equation of state by a perturbation method. i.
  nonpolar gases}.
\bjtitle{The Journal of Chemical Physics}
\bvolume{22}(\bissue{8}),
\bfpage{1420}--\blpage{1426}
(\byear{1954})
\doiurl{10.1063/1.1740409}
\end{barticle}
\endbibitem

\bibitem[\protect\citeauthoryear{Bennett}{1976}]{Bennett_bar_1976}
\begin{barticle}
\bauthor{\bsnm{Bennett}, \binits{C.H.}}:
\batitle{Efficient estimation of free energy differences from monte carlo
  data}.
\bjtitle{Journal of Computational Physics}
\bvolume{22}(\bissue{2}),
\bfpage{245}--\blpage{268}
(\byear{1976})
\doiurl{10.1016/0021-9991(76)90078-4}
\end{barticle}
\endbibitem

\bibitem[\protect\citeauthoryear{Lahey et~al.}{2020}]{lahey_2020_biaryl}
\begin{barticle}
\bauthor{\bsnm{Lahey}, \binits{S.-L.J.}},
\bauthor{\bsnm{Thien~Phuc}, \binits{T.N.}},
\bauthor{\bsnm{Rowley}, \binits{C.N.}}:
\batitle{Benchmarking force field and the ani neural network potentials for the
  torsional potential energy surface of biaryl drug fragments}.
\bjtitle{Journal of Chemical Information and Modeling}
\bvolume{60}(\bissue{12}),
\bfpage{6258}--\blpage{6268}
(\byear{2020})
\doiurl{10.1021/acs.jcim.0c00904}
{\href{https://arxiv.org/abs/https://doi.org/10.1021/acs.jcim.0c00904}{{https://doi.org/10.1021/acs.jcim.0c00904}}}.
\bcomment{PMID: 33263401}
\end{barticle}
\endbibitem

\bibitem[\protect\citeauthoryear{Waibl
  et~al.}{2024}]{noeMacrocyclicCompounds_Waibl_2024}
\begin{barticle}
\bauthor{\bsnm{Waibl}, \binits{F.}},
\bauthor{\bsnm{Casagrande}, \binits{F.}},
\bauthor{\bsnm{Dey}, \binits{F.}},
\bauthor{\bsnm{Riniker}, \binits{S.}}:
\batitle{Validating small-molecule force fields for macrocyclic compounds using
  nmr data in different solvents}.
\bjtitle{Journal of Chemical Information and Modeling}
\bvolume{64}(\bissue{20}),
\bfpage{7938}--\blpage{7948}
(\byear{2024})
\doiurl{10.1021/acs.jcim.4c01120}
{\href{https://arxiv.org/abs/https://doi.org/10.1021/acs.jcim.4c01120}{{https://doi.org/10.1021/acs.jcim.4c01120}}}.
\bcomment{PMID: 39405498}
\end{barticle}
\endbibitem

\bibitem[\protect\citeauthoryear{Cohen et~al.}{2016}]{ozanimod_Cohen_2016}
\begin{barticle}
\bauthor{\bsnm{Cohen}, \binits{J.A.}},
\bauthor{\bsnm{Arnold}, \binits{D.L.}},
\bauthor{\bsnm{Comi}, \binits{G.}},
\bauthor{\bsnm{Bar-Or}, \binits{A.}},
\bauthor{\bsnm{Gujrathi}, \binits{S.}},
\bauthor{\bsnm{Hartung}, \binits{J.P.}},
\bauthor{\bsnm{Cravets}, \binits{M.}},
\bauthor{\bsnm{Olson}, \binits{A.}},
\bauthor{\bsnm{Frohna}, \binits{P.A.}},
\bauthor{\bsnm{Selmaj}, \binits{K.W.}}:
\batitle{Safety and efficacy of the selective sphingosine 1-phosphate receptor
  modulator ozanimod in relapsing multiple sclerosis (radiance): a randomised,
  placebo-controlled, phase 2 trial}.
\bjtitle{The Lancet Neurology}
\bvolume{15}(\bissue{4}),
\bfpage{373}--\blpage{381}
(\byear{2016})
\doiurl{10.1016/S1474-4422(16)00018-1}
\end{barticle}
\endbibitem

\bibitem[\protect\citeauthoryear{Driggers
  et~al.}{2008}]{macrocyclesfordrugdiscovery_Driggers_2008}
\begin{barticle}
\bauthor{\bsnm{Driggers}, \binits{E.M.}},
\bauthor{\bsnm{Hale}, \binits{S.P.}},
\bauthor{\bsnm{Lee}, \binits{J.}},
\bauthor{\bsnm{Terrett}, \binits{N.K.}}:
\batitle{The exploration of macrocycles for drug discovery — an
  underexploited structural class}.
\bjtitle{Nature Reviews Drug Discovery}
\bvolume{7}(\bissue{7}),
\bfpage{608}--\blpage{624}
(\byear{2008})
\doiurl{10.1038/nrd2590} .
Accessed 2026-02-25
\end{barticle}
\endbibitem

\bibitem[\protect\citeauthoryear{Sethio
  et~al.}{2023}]{SimulationReveals_Sethio_2023}
\begin{barticle}
\bauthor{\bsnm{Sethio}, \binits{D.}},
\bauthor{\bsnm{Poongavanam}, \binits{V.}},
\bauthor{\bsnm{Xiong}, \binits{R.}},
\bauthor{\bsnm{Tyagi}, \binits{M.}},
\bauthor{\bsnm{Duy~Vo}, \binits{D.}},
\bauthor{\bsnm{Lindh}, \binits{R.}},
\bauthor{\bsnm{Kihlberg}, \binits{J.}}:
\batitle{Simulation reveals the chameleonic behavior of macrocycles}.
\bjtitle{Journal of Chemical Information and Modeling}
\bvolume{63}(\bissue{1}),
\bfpage{138}--\blpage{146}
(\byear{2023})
\doiurl{10.1021/acs.jcim.2c01093}
{\href{https://arxiv.org/abs/https://doi.org/10.1021/acs.jcim.2c01093}{{https://doi.org/10.1021/acs.jcim.2c01093}}}.
\bcomment{PMID: 36563083}
\end{barticle}
\endbibitem

\bibitem[\protect\citeauthoryear{Kamenik
  et~al.}{2018}]{PeptidicMacrocycles_Kamenik_2028}
\begin{barticle}
\bauthor{\bsnm{Kamenik}, \binits{A.S.}},
\bauthor{\bsnm{Lessel}, \binits{U.}},
\bauthor{\bsnm{Fuchs}, \binits{J.E.}},
\bauthor{\bsnm{Fox}, \binits{T.}},
\bauthor{\bsnm{Liedl}, \binits{K.R.}}:
\batitle{Peptidic macrocycles - conformational sampling and thermodynamic
  characterization}.
\bjtitle{Journal of Chemical Information and Modeling}
\bvolume{58}(\bissue{5}),
\bfpage{982}--\blpage{992}
(\byear{2018})
\doiurl{10.1021/acs.jcim.8b00097}
{\href{https://arxiv.org/abs/https://doi.org/10.1021/acs.jcim.8b00097}{{https://doi.org/10.1021/acs.jcim.8b00097}}}.
\bcomment{PMID: 29652495}
\end{barticle}
\endbibitem

\bibitem[\protect\citeauthoryear{Zhang
  et~al.}{2020}]{Zhang_alanineInWater_2020}
\begin{barticle}
\bauthor{\bsnm{Zhang}, \binits{S.}},
\bauthor{\bsnm{Schweitzer-Stenner}, \binits{R.}},
\bauthor{\bsnm{Urbanc}, \binits{B.}}:
\batitle{Do molecular dynamics force fields capture conformational dynamics of
  alanine in water?}
\bjtitle{Journal of Chemical Theory and Computation}
\bvolume{16}(\bissue{1}),
\bfpage{510}--\blpage{527}
(\byear{2020})
\doiurl{10.1021/acs.jctc.9b00588}
{\href{https://arxiv.org/abs/https://doi.org/10.1021/acs.jctc.9b00588}{{https://doi.org/10.1021/acs.jctc.9b00588}}}.
\bcomment{PMID: 31751129}
\end{barticle}
\endbibitem

\bibitem[\protect\citeauthoryear{Esadze
  et~al.}{2011}]{DynamicsLysineSide_Esadze_2011}
\begin{barticle}
\bauthor{\bsnm{Esadze}, \binits{A.}},
\bauthor{\bsnm{Li}, \binits{D.-W.}},
\bauthor{\bsnm{Wang}, \binits{T.}},
\bauthor{\bsnm{Br{\"u}schweiler}, \binits{R.}},
\bauthor{\bsnm{Iwahara}, \binits{J.}}:
\batitle{Dynamics of lysine side-chain amino groups in a protein studied by
  heteronuclear 1h-15n nmr spectroscopy}.
\bjtitle{Journal of the American Chemical Society}
\bvolume{133}(\bissue{4}),
\bfpage{909}--\blpage{919}
(\byear{2011})
\doiurl{10.1021/ja107847d}
{\href{https://arxiv.org/abs/https://doi.org/10.1021/ja107847d}{{https://doi.org/10.1021/ja107847d}}}.
\bcomment{PMID: 21186799}
\end{barticle}
\endbibitem

\bibitem[\protect\citeauthoryear{Rosenberger
  et~al.}{2021}]{Rosenberger_peptidesmlp_2021}
\begin{barticle}
\bauthor{\bsnm{Rosenberger}, \binits{D.}},
\bauthor{\bsnm{Smith}, \binits{J.S.}},
\bauthor{\bsnm{Garcia}, \binits{A.E.}}:
\batitle{Modeling of peptides with classical and novel machine learning force
  fields: A comparison}.
\bjtitle{The Journal of Physical Chemistry B}
\bvolume{125}(\bissue{14}),
\bfpage{3598}--\blpage{3612}
(\byear{2021})
\doiurl{10.1021/acs.jpcb.0c10401}
{\href{https://arxiv.org/abs/https://doi.org/10.1021/acs.jpcb.0c10401}{{https://doi.org/10.1021/acs.jpcb.0c10401}}}.
\bcomment{PMID: 33798336}
\end{barticle}
\endbibitem

\bibitem[\protect\citeauthoryear{Engel et~al.}{2021}]{Engel_importanceNQE_2021}
\begin{barticle}
\bauthor{\bsnm{Engel}, \binits{E.A.}},
\bauthor{\bsnm{Kapil}, \binits{V.}},
\bauthor{\bsnm{Ceriotti}, \binits{M.}}:
\batitle{Importance of nuclear quantum effects for nmr crystallography}.
\bjtitle{The Journal of Physical Chemistry Letters}
\bvolume{12}(\bissue{32}),
\bfpage{7701}--\blpage{7707}
(\byear{2021})
\doiurl{10.1021/acs.jpclett.1c01987}
{\href{https://arxiv.org/abs/https://doi.org/10.1021/acs.jpclett.1c01987}{{https://doi.org/10.1021/acs.jpclett.1c01987}}}.
\bcomment{PMID: 34355903}
\end{barticle}
\endbibitem

\bibitem[\protect\citeauthoryear{Huang and
  MacKerell~Jr}{2013}]{huang_2013_CHARMM36nmr}
\begin{barticle}
\bauthor{\bsnm{Huang}, \binits{J.}},
\bauthor{\bsnm{MacKerell~Jr}, \binits{A.D.}}:
\batitle{Charmm36 all-atom additive protein force field: Validation based on
  comparison to nmr data}.
\bjtitle{Journal of Computational Chemistry}
\bvolume{34}(\bissue{25}),
\bfpage{2135}--\blpage{2145}
(\byear{2013})
\doiurl{10.1002/jcc.23354}
{\href{https://arxiv.org/abs/https://onlinelibrary.wiley.com/doi/pdf/10.1002/jcc.23354}{{https://onlinelibrary.wiley.com/doi/pdf/10.1002/jcc.23354}}}
\end{barticle}
\endbibitem

\bibitem[\protect\citeauthoryear{Lange
  et~al.}{2010}]{ScrutinizingMolecularMechanics_Lange_2010}
\begin{barticle}
\bauthor{\bsnm{Lange}, \binits{O.F.}},
\bauthor{\bsnm{{van der Spoel}}, \binits{D.}},
\bauthor{\bsnm{{de Groot}}, \binits{B.L.}}:
\batitle{Scrutinizing molecular mechanics force fields on the submicrosecond
  timescale with nmr data}.
\bjtitle{Biophysical Journal}
\bvolume{99}(\bissue{2}),
\bfpage{647}--\blpage{655}
(\byear{2010})
\doiurl{10.1016/j.bpj.2010.04.062}
\end{barticle}
\endbibitem

\bibitem[\protect\citeauthoryear{Lindorff-Larsen
  et~al.}{2012}]{SystematicValidation_LindorffLarsen_2012}
\begin{barticle}
\bauthor{\bsnm{Lindorff-Larsen}, \binits{K.}},
\bauthor{\bsnm{Maragakis}, \binits{P.}},
\bauthor{\bsnm{Piana}, \binits{S.}},
\bauthor{\bsnm{Eastwood}, \binits{M.P.}},
\bauthor{\bsnm{Dror}, \binits{R.O.}},
\bauthor{\bsnm{Shaw}, \binits{D.E.}}:
\batitle{Systematic validation of protein force fields against experimental
  data}.
\bjtitle{PLOS ONE}
\bvolume{7}(\bissue{2}),
\bfpage{1}--\blpage{6}
(\byear{2012})
\doiurl{10.1371/journal.pone.0032131}
\end{barticle}
\endbibitem

\bibitem[\protect\citeauthoryear{Beauchamp
  et~al.}{2012}]{AreProteinForce_Beauchamp_2012}
\begin{barticle}
\bauthor{\bsnm{Beauchamp}, \binits{K.A.}},
\bauthor{\bsnm{Lin}, \binits{Y.-S.}},
\bauthor{\bsnm{Das}, \binits{R.}},
\bauthor{\bsnm{Pande}, \binits{V.S.}}:
\batitle{Are protein force fields getting better? a systematic benchmark on 524
  diverse nmr measurements}.
\bjtitle{Journal of Chemical Theory and Computation}
\bvolume{8}(\bissue{4}),
\bfpage{1409}--\blpage{1414}
(\byear{2012})
\doiurl{10.1021/ct2007814}
{\href{https://arxiv.org/abs/https://doi.org/10.1021/ct2007814}{{https://doi.org/10.1021/ct2007814}}}.
\bcomment{PMID: 22754404}
\end{barticle}
\endbibitem

\bibitem[\protect\citeauthoryear{Juranić
  et~al.}{2002}]{juranic_h3nc_1ubq_2002}
\begin{barticle}
\bauthor{\bsnm{Juranić}, \binits{N.}},
\bauthor{\bsnm{Moncrieffe}, \binits{M.C.}},
\bauthor{\bsnm{Likić}, \binits{V.A.}},
\bauthor{\bsnm{Prendergast}, \binits{F.G.}},
\bauthor{\bsnm{Macura}, \binits{S.}}:
\batitle{Structural dependencies of h3jnc‘ scalar coupling in protein h-bond
  chains}.
\bjtitle{Journal of the American Chemical Society}
\bvolume{124}(\bissue{47}),
\bfpage{14221}--\blpage{14226}
(\byear{2002})
\doiurl{10.1021/ja0273288}
{\href{https://arxiv.org/abs/https://doi.org/10.1021/ja0273288}{{https://doi.org/10.1021/ja0273288}}}.
\bcomment{PMID: 12440921}
\end{barticle}
\endbibitem

\bibitem[\protect\citeauthoryear{Alexandrescu
  et~al.}{2001}]{alexandrescu_h3nc_1mjc_2001}
\begin{barticle}
\bauthor{\bsnm{Alexandrescu}, \binits{A.T.}},
\bauthor{\bsnm{Snyder}, \binits{D.R.}},
\bauthor{\bsnm{Abildgaard}, \binits{F.}}:
\batitle{Nmr of hydrogen bonding in cold-shock protein a and an analysis of the
  influence of crystallographic resolution on comparisons of hydrogen bond
  lengths}.
\bjtitle{Protein Science}
\bvolume{10}(\bissue{9}),
\bfpage{1856}--\blpage{1868}
(\byear{2001})
\doiurl{10.1110/ps.14301}
{\href{https://arxiv.org/abs/https://onlinelibrary.wiley.com/doi/pdf/10.1110/ps.14301}{{https://onlinelibrary.wiley.com/doi/pdf/10.1110/ps.14301}}}
\end{barticle}
\endbibitem

\bibitem[\protect\citeauthoryear{Cornilescu
  et~al.}{1999}]{Cornilescu_h3jnc_2qmt_1999}
\begin{barticle}
\bauthor{\bsnm{Cornilescu}, \binits{G.}},
\bauthor{\bsnm{Ramirez}, \binits{B.E.}},
\bauthor{\bsnm{Frank}, \binits{M.K.}},
\bauthor{\bsnm{Clore}, \binits{G.M.}},
\bauthor{\bsnm{Gronenborn}, \binits{A.M.}},
\bauthor{\bsnm{Bax}, \binits{A.}}:
\batitle{Correlation between 3hjnc‘ and hydrogen bond length in proteins}.
\bjtitle{Journal of the American Chemical Society}
\bvolume{121}(\bissue{26}),
\bfpage{6275}--\blpage{6279}
(\byear{1999})
\doiurl{10.1021/ja9909024}
{\href{https://arxiv.org/abs/https://doi.org/10.1021/ja9909024}{{https://doi.org/10.1021/ja9909024}}}
\end{barticle}
\endbibitem

\bibitem[\protect\citeauthoryear{Tjandra
  et~al.}{1995}]{Tjandra_15NNMRrelaxation_1995}
\begin{barticle}
\bauthor{\bsnm{Tjandra}, \binits{N.}},
\bauthor{\bsnm{Feller}, \binits{S.E.}},
\bauthor{\bsnm{Pastor}, \binits{R.W.}},
\bauthor{\bsnm{Bax}, \binits{A.}}:
\batitle{Rotational diffusion anisotropy of human ubiquitin from 15n nmr
  relaxation}.
\bjtitle{Journal of the American Chemical Society}
\bvolume{117}(\bissue{50}),
\bfpage{12562}--\blpage{12566}
(\byear{1995})
\doiurl{10.1021/ja00155a020}
{\href{https://arxiv.org/abs/https://doi.org/10.1021/ja00155a020}{{https://doi.org/10.1021/ja00155a020}}}
\end{barticle}
\endbibitem

\bibitem[\protect\citeauthoryear{Esadze et~al.}{2011}]{esadze_j3cn_1ubq_2011}
\begin{barticle}
\bauthor{\bsnm{Esadze}, \binits{A.}},
\bauthor{\bsnm{Li}, \binits{D.-W.}},
\bauthor{\bsnm{Wang}, \binits{T.}},
\bauthor{\bsnm{Brüschweiler}, \binits{R.}},
\bauthor{\bsnm{Iwahara}, \binits{J.}}:
\batitle{Dynamics of lysine side-chain amino groups in a protein studied by
  heteronuclear 1h-15n nmr spectroscopy}.
\bjtitle{Journal of the American Chemical Society}
\bvolume{133}(\bissue{4}),
\bfpage{909}--\blpage{919}
(\byear{2011})
\doiurl{10.1021/ja107847d}
{\href{https://arxiv.org/abs/https://doi.org/10.1021/ja107847d}{{https://doi.org/10.1021/ja107847d}}}.
\bcomment{PMID: 21186799}
\end{barticle}
\endbibitem

\bibitem[\protect\citeauthoryear{Lee et~al.}{1999}]{lee_s2_methyl_1ubq_1999}
\begin{barticle}
\bauthor{\bsnm{Lee}, \binits{A.L.}},
\bauthor{\bsnm{Flynn}, \binits{P.F.}},
\bauthor{\bsnm{Wand}, \binits{A.J.}}:
\batitle{Comparison of 2h and 13c nmr relaxation techniques for the study of
  protein methyl group dynamics in solution}.
\bjtitle{Journal of the American Chemical Society}
\bvolume{121}(\bissue{12}),
\bfpage{2891}--\blpage{2902}
(\byear{1999})
\doiurl{10.1021/ja983758f}
{\href{https://arxiv.org/abs/https://doi.org/10.1021/ja983758f}{{https://doi.org/10.1021/ja983758f}}}
\end{barticle}
\endbibitem

\bibitem[\protect\citeauthoryear{Kim
  et~al.}{2025}]{AUniversalAugmentationFramework_Kim_2025}
\begin{barticle}
\bauthor{\bsnm{Kim}, \binits{D.}},
\bauthor{\bsnm{Wang}, \binits{X.}},
\bauthor{\bsnm{Vargas}, \binits{S.}},
\bauthor{\bsnm{Zhong}, \binits{P.}},
\bauthor{\bsnm{King}, \binits{D.S.}},
\bauthor{\bsnm{Inizan}, \binits{T.J.}},
\bauthor{\bsnm{Cheng}, \binits{B.}}:
\batitle{A universal augmentation framework for long-range electrostatics in
  machine learning interatomic potentials}.
\bjtitle{Journal of Chemical Theory and Computation}
\bvolume{21}(\bissue{24}),
\bfpage{12709}--\blpage{12724}
(\byear{2025})
\doiurl{10.1021/acs.jctc.5c01400}
{\href{https://arxiv.org/abs/https://doi.org/10.1021/acs.jctc.5c01400}{{https://doi.org/10.1021/acs.jctc.5c01400}}}.
\bcomment{PMID: 41368735}
\end{barticle}
\endbibitem

\bibitem[\protect\citeauthoryear{Henriques
  et~al.}{2015}]{Henriques_mdDisorderedProteins_2015}
\begin{barticle}
\bauthor{\bsnm{Henriques}, \binits{J.}},
\bauthor{\bsnm{Cragnell}, \binits{C.}},
\bauthor{\bsnm{Skep{\"o}}, \binits{M.}}:
\batitle{Molecular dynamics simulations of intrinsically disordered proteins:
  Force field evaluation and comparison with experiment}.
\bjtitle{Journal of Chemical Theory and Computation}
\bvolume{11}(\bissue{7}),
\bfpage{3420}--\blpage{3431}
(\byear{2015})
\doiurl{10.1021/ct501178z}
{\href{https://arxiv.org/abs/https://doi.org/10.1021/ct501178z}{{https://doi.org/10.1021/ct501178z}}}.
\bcomment{PMID: 26575776}
\end{barticle}
\endbibitem

\bibitem[\protect\citeauthoryear{Huang and
  MacKerell}{2018}]{Huang_ffdisorderedproteins_2018}
\begin{barticle}
\bauthor{\bsnm{Huang}, \binits{J.}},
\bauthor{\bsnm{MacKerell}, \binits{A.D.}}:
\batitle{Force field development and simulations of intrinsically disordered
  proteins}.
\bjtitle{Current Opinion in Structural Biology}
\bvolume{48},
\bfpage{40}--\blpage{48}
(\byear{2018})
\doiurl{10.1016/j.sbi.2017.10.008} .
\bcomment{Folding and binding in silico, in vitro and in cellula • Proteins:
  An Evolutionary Perspective}
\end{barticle}
\endbibitem

\bibitem[\protect\citeauthoryear{Rauscher
  et~al.}{2015}]{Rauscher_StructuralEnsemblesIDP_2015}
\begin{barticle}
\bauthor{\bsnm{Rauscher}, \binits{S.}},
\bauthor{\bsnm{Gapsys}, \binits{V.}},
\bauthor{\bsnm{Gajda}, \binits{M.J.}},
\bauthor{\bsnm{Zweckstetter}, \binits{M.}},
\bauthor{\bsnm{Groot}, \binits{B.L.}},
\bauthor{\bsnm{Grubm{\"u}ller}, \binits{H.}}:
\batitle{Structural ensembles of intrinsically disordered proteins depend
  strongly on force field: A comparison to experiment}.
\bjtitle{Journal of Chemical Theory and Computation}
\bvolume{11}(\bissue{11}),
\bfpage{5513}--\blpage{5524}
(\byear{2015})
\doiurl{10.1021/acs.jctc.5b00736}
{\href{https://arxiv.org/abs/https://doi.org/10.1021/acs.jctc.5b00736}{{https://doi.org/10.1021/acs.jctc.5b00736}}}.
\bcomment{PMID: 26574339}
\end{barticle}
\endbibitem

\bibitem[\protect\citeauthoryear{Husic et~al.}{2020}]{Husic_cgmd_2020}
\begin{barticle}
\bauthor{\bsnm{Husic}, \binits{B.E.}},
\bauthor{\bsnm{Charron}, \binits{N.E.}},
\bauthor{\bsnm{Lemm}, \binits{D.}},
\bauthor{\bsnm{Wang}, \binits{J.}},
\bauthor{\bsnm{Pérez}, \binits{A.}},
\bauthor{\bsnm{Majewski}, \binits{M.}},
\bauthor{\bsnm{Krämer}, \binits{A.}},
\bauthor{\bsnm{Chen}, \binits{Y.}},
\bauthor{\bsnm{Olsson}, \binits{S.}},
\bauthor{\bsnm{Fabritiis}, \binits{G.}},
\bauthor{\bsnm{Noé}, \binits{F.}},
\bauthor{\bsnm{Clementi}, \binits{C.}}:
\batitle{Coarse graining molecular dynamics with graph neural networks}.
\bjtitle{The Journal of Chemical Physics}
\bvolume{153}(\bissue{19}),
\bfpage{194101}
(\byear{2020})
\doiurl{10.1063/5.0026133}
\end{barticle}
\endbibitem

\bibitem[\protect\citeauthoryear{Noid}{2013}]{Noid_cgmodelsBio_2013}
\begin{barticle}
\bauthor{\bsnm{Noid}, \binits{W.G.}}:
\batitle{Perspective: Coarse-grained models for biomolecular systems}.
\bjtitle{The Journal of Chemical Physics}
\bvolume{139}(\bissue{9}),
\bfpage{090901}
(\byear{2013})
\doiurl{10.1063/1.4818908}
\end{barticle}
\endbibitem

\bibitem[\protect\citeauthoryear{Kar and
  Feig}{2014}]{Kar_AdvancesinProteinChemistry_2014}
\begin{bchapter}
\bauthor{\bsnm{Kar}, \binits{P.}},
\bauthor{\bsnm{Feig}, \binits{M.}}:
\bctitle{Chapter five - recent advances in transferable coarse-grained modeling
  of proteins}.
In: \beditor{\bsnm{Karabencheva-Christova}, \binits{T.}} (ed.)
\bbtitle{Biomolecular Modelling and Simulations}.
\bsertitle{Advances in Protein Chemistry and Structural Biology},
vol. \bseriesno{96},
pp. \bfpage{143}--\blpage{180}.
\bpublisher{Academic Press}, \blocation{???}
(\byear{2014}).
\doiurl{10.1016/bs.apcsb.2014.06.005} .
\burl{https://www.sciencedirect.com/science/article/pii/S1876162314000066}
\end{bchapter}
\endbibitem

\bibitem[\protect\citeauthoryear{Cornilescu
  et~al.}{1999}]{Correlationbetween_Cornilescu_1999}
\begin{barticle}
\bauthor{\bsnm{Cornilescu}, \binits{G.}},
\bauthor{\bsnm{Ramirez}, \binits{B.E.}},
\bauthor{\bsnm{Frank}, \binits{M.K.}},
\bauthor{\bsnm{Clore}, \binits{G.M.}},
\bauthor{\bsnm{Gronenborn}, \binits{A.M.}},
\bauthor{\bsnm{Bax}, \binits{A.}}:
\batitle{Correlation between 3hjnc‘ and hydrogen bond length in proteins}.
\bjtitle{Journal of the American Chemical Society}
\bvolume{121}(\bissue{26}),
\bfpage{6275}--\blpage{6279}
(\byear{1999})
\doiurl{10.1021/ja9909024}
{\href{https://arxiv.org/abs/https://doi.org/10.1021/ja9909024}{{https://doi.org/10.1021/ja9909024}}}
\end{barticle}
\endbibitem

\bibitem[\protect\citeauthoryear{Grzesiek
  et~al.}{2004}]{Insightsinto_Grzesiek_2004}
\begin{barticle}
\bauthor{\bsnm{Grzesiek}, \binits{S.}},
\bauthor{\bsnm{Cordier}, \binits{F.}},
\bauthor{\bsnm{Jaravine}, \binits{V.}},
\bauthor{\bsnm{Barfield}, \binits{M.}}:
\batitle{Insights into biomolecular hydrogen bonds from hydrogen bond scalar
  couplings}.
\bjtitle{Progress in Nuclear Magnetic Resonance Spectroscopy}
\bvolume{45}(\bissue{3}),
\bfpage{275}--\blpage{300}
(\byear{2004})
\doiurl{10.1016/j.pnmrs.2004.08.001}
\end{barticle}
\endbibitem

\bibitem[\protect\citeauthoryear{Barfield}{2002}]{StructuralDependencies_Barfield_2002}
\begin{barticle}
\bauthor{\bsnm{Barfield}, \binits{M.}}:
\batitle{Structural dependencies of interresidue scalar coupling h3jnc‘ and
  donor 1h chemical shifts in the hydrogen bonding regions of proteins}.
\bjtitle{Journal of the American Chemical Society}
\bvolume{124}(\bissue{15}),
\bfpage{4158}--\blpage{4168}
(\byear{2002})
\doiurl{10.1021/ja012674v}
{\href{https://arxiv.org/abs/https://doi.org/10.1021/ja012674v}{{https://doi.org/10.1021/ja012674v}}}.
\bcomment{PMID: 11942855}
\end{barticle}
\endbibitem

\bibitem[\protect\citeauthoryear{Nisius and
  Grzesiek}{2012}]{Keystabilizingelements_Nisius_2012}
\begin{barticle}
\bauthor{\bsnm{Nisius}, \binits{L.}},
\bauthor{\bsnm{Grzesiek}, \binits{S.}}:
\batitle{Key stabilizing elements of protein structure identified through
  pressure and temperature perturbation of its hydrogen bond network}.
\bjtitle{Nature Chem}
\bvolume{4},
\bfpage{711}--\blpage{717}
(\byear{2012})
\doiurl{10.1038/nchem.1396}
\end{barticle}
\endbibitem

\bibitem[\protect\citeauthoryear{Zhou and Wang}{2019}]{Zhou_shorthbonds_2019}
\begin{barticle}
\bauthor{\bsnm{Zhou}, \binits{S.}},
\bauthor{\bsnm{Wang}, \binits{L.}}:
\batitle{Unraveling the structural and chemical features of biological short
  hydrogen bonds}.
\bjtitle{Chemical Science}
\bvolume{10}(\bissue{33}),
\bfpage{7734}--\blpage{7745}
(\byear{2019})
\doiurl{10.1039/c9sc01496a}
\end{barticle}
\endbibitem

\bibitem[\protect\citeauthoryear{Ogata et~al.}{2013}]{Ogata_nqe_2013}
\begin{barticle}
\bauthor{\bsnm{Ogata}, \binits{Y.}},
\bauthor{\bsnm{Daido}, \binits{M.}},
\bauthor{\bsnm{Kawashima}, \binits{Y.}},
\bauthor{\bsnm{Tachikawa}, \binits{M.}}:
\batitle{Nuclear quantum effects on protonated lysine with an asymmetric low
  barrier hydrogen bond: an ab initio path integral molecular dynamics study}.
\bjtitle{RSC Adv.}
\bvolume{3},
\bfpage{25252}--\blpage{25257}
(\byear{2013})
\doiurl{10.1039/C3RA44077J}
\end{barticle}
\endbibitem

\bibitem[\protect\citeauthoryear{Cornilescu
  et~al.}{1999}]{Cornilescu_IdentificationHBN_1999}
\begin{barticle}
\bauthor{\bsnm{Cornilescu}, \binits{G.}},
\bauthor{\bsnm{Hu}, \binits{J.-S.}},
\bauthor{\bsnm{Bax}, \binits{A.}}:
\batitle{Identification of the hydrogen bonding network in a protein by scalar
  couplings}.
\bjtitle{Journal of the American Chemical Society}
\bvolume{121}(\bissue{12}),
\bfpage{2949}--\blpage{2950}
(\byear{1999})
\doiurl{10.1021/ja9902221}
{\href{https://arxiv.org/abs/https://doi.org/10.1021/ja9902221}{{https://doi.org/10.1021/ja9902221}}}
\end{barticle}
\endbibitem

\bibitem[\protect\citeauthoryear{Zandarashvili
  et~al.}{2011}]{SignatureofMobile_Zandarashvili_2011}
\begin{barticle}
\bauthor{\bsnm{Zandarashvili}, \binits{L.}},
\bauthor{\bsnm{Li}, \binits{D.-W.}},
\bauthor{\bsnm{Wang}, \binits{T.}},
\bauthor{\bsnm{Br{\"u}schweiler}, \binits{R.}},
\bauthor{\bsnm{Iwahara}, \binits{J.}}:
\batitle{Signature of mobile hydrogen bonding of lysine side chains from
  long-range 15n–13c scalar j-couplings and computation}.
\bjtitle{Journal of the American Chemical Society}
\bvolume{133}(\bissue{24}),
\bfpage{9192}--\blpage{9195}
(\byear{2011})
\doiurl{10.1021/ja202219n}
{\href{https://arxiv.org/abs/https://doi.org/10.1021/ja202219n}{{https://doi.org/10.1021/ja202219n}}}.
\bcomment{PMID: 21591797}
\end{barticle}
\endbibitem

\bibitem[\protect\citeauthoryear{Kříž and
  Řezáč}{2020}]{Kristian_pla15_2020}
\begin{barticle}
\bauthor{\bsnm{Kříž}, \binits{K.}},
\bauthor{\bsnm{Řezáč}, \binits{J.}}:
\batitle{Benchmarking of semiempirical quantum-mechanical methods on systems
  relevant to computer-aided drug design}.
\bjtitle{Journal of Chemical Information and Modeling}
\bvolume{60}(\bissue{3}),
\bfpage{1453}--\blpage{1460}
(\byear{2020})
\doiurl{10.1021/acs.jcim.9b01171}
{\href{https://arxiv.org/abs/https://doi.org/10.1021/acs.jcim.9b01171}{{https://doi.org/10.1021/acs.jcim.9b01171}}}.
\bcomment{PMID: 32062970}
\end{barticle}
\endbibitem

\bibitem[\protect\citeauthoryear{Batatia et~al.}{2026}]{batatia_macepolar_2026}
\begin{botherref}
\oauthor{\bsnm{Batatia}, \binits{I.}},
\oauthor{\bsnm{Baldwin}, \binits{W.J.}},
\oauthor{\bsnm{Kuryla}, \binits{D.}},
\oauthor{\bsnm{Hart}, \binits{J.}},
\oauthor{\bsnm{Kasoar}, \binits{E.}},
\oauthor{\bsnm{Elena}, \binits{A.M.}},
\oauthor{\bsnm{Moore}, \binits{H.}},
\oauthor{\bsnm{Gawkowski}, \binits{M.J.}},
\oauthor{\bsnm{Shi}, \binits{B.X.}},
\oauthor{\bsnm{Kapil}, \binits{V.}},
\oauthor{\bsnm{Kourtis}, \binits{P.}},
\oauthor{\bsnm{Magdău}, \binits{I.-B.}},
\oauthor{\bsnm{Csányi}, \binits{G.}}:
MACE-POLAR-1: A Polarisable Electrostatic Foundation Model for Molecular
  Chemistry
(2026).
\url{https://arxiv.org/abs/2602.19411}
\end{botherref}
\endbibitem

\bibitem[\protect\citeauthoryear{Thompson et~al.}{2022}]{Thompson_lammps_2022}
\begin{barticle}
\bauthor{\bsnm{Thompson}, \binits{A.P.}},
\bauthor{\bsnm{Aktulga}, \binits{H.M.}},
\bauthor{\bsnm{Berger}, \binits{R.}},
\bauthor{\bsnm{Bolintineanu}, \binits{D.S.}},
\bauthor{\bsnm{Brown}, \binits{W.M.}},
\bauthor{\bsnm{Crozier}, \binits{P.S.}},
\bauthor{\bsnm{{in 't Veld}}, \binits{P.J.}},
\bauthor{\bsnm{Kohlmeyer}, \binits{A.}},
\bauthor{\bsnm{Moore}, \binits{S.G.}},
\bauthor{\bsnm{Nguyen}, \binits{T.D.}},
\bauthor{\bsnm{Shan}, \binits{R.}},
\bauthor{\bsnm{Stevens}, \binits{M.J.}},
\bauthor{\bsnm{Tranchida}, \binits{J.}},
\bauthor{\bsnm{Trott}, \binits{C.}},
\bauthor{\bsnm{Plimpton}, \binits{S.J.}}:
\batitle{Lammps - a flexible simulation tool for particle-based materials
  modeling at the atomic, meso, and continuum scales}.
\bjtitle{Computer Physics Communications}
\bvolume{271},
\bfpage{108171}
(\byear{2022})
\doiurl{10.1016/j.cpc.2021.108171}
\end{barticle}
\endbibitem

\bibitem[\protect\citeauthoryear{Bharadwaj
  et~al.}{2025}]{Bharadwaj_openequivariance_2025}
\begin{bbook}
\bauthor{\bsnm{Bharadwaj}, \binits{V.}},
\bauthor{\bsnm{Glover}, \binits{A.}},
\bauthor{\bsnm{Buluc}, \binits{A.}},
\bauthor{\bsnm{Demmel}, \binits{J.}}:
\bbtitle{An Efficient Sparse Kernel Generator for O(3)-Equivariant Deep
  Networks}.
\bpublisher{Society for Industrial and Applied Mathematics}, \blocation{???}
(\byear{2025}).
\burl{https://arxiv.org/abs/2501.13986}
\end{bbook}
\endbibitem

\bibitem[\protect\citeauthoryear{Eastman et~al.}{2024}]{Eastman_OpenMM_2024}
\begin{barticle}
\bauthor{\bsnm{Eastman}, \binits{P.}},
\bauthor{\bsnm{Galvelis}, \binits{R.}},
\bauthor{\bsnm{Peláez}, \binits{R.P.}},
\bauthor{\bsnm{Abreu}, \binits{C.R.A.}},
\bauthor{\bsnm{Farr}, \binits{S.E.}},
\bauthor{\bsnm{Gallicchio}, \binits{E.}},
\bauthor{\bsnm{Gorenko}, \binits{A.}},
\bauthor{\bsnm{Henry}, \binits{M.M.}},
\bauthor{\bsnm{Hu}, \binits{F.}},
\bauthor{\bsnm{Huang}, \binits{J.}},
\bauthor{\bsnm{Krämer}, \binits{A.}},
\bauthor{\bsnm{Michel}, \binits{J.}},
\bauthor{\bsnm{Mitchell}, \binits{J.A.}},
\bauthor{\bsnm{Pande}, \binits{V.S.}},
\bauthor{\bsnm{Rodrigues}, \binits{J.P.}},
\bauthor{\bsnm{Rodriguez-Guerra}, \binits{J.}},
\bauthor{\bsnm{Simmonett}, \binits{A.C.}},
\bauthor{\bsnm{Singh}, \binits{S.}},
\bauthor{\bsnm{Swails}, \binits{J.}},
\bauthor{\bsnm{Turner}, \binits{P.}},
\bauthor{\bsnm{Wang}, \binits{Y.}},
\bauthor{\bsnm{Zhang}, \binits{I.}},
\bauthor{\bsnm{Chodera}, \binits{J.D.}},
\bauthor{\bsnm{De~Fabritiis}, \binits{G.}},
\bauthor{\bsnm{Markland}, \binits{T.E.}}:
\batitle{Openmm 8: Molecular dynamics simulation with machine learning
  potentials}.
\bjtitle{The Journal of Physical Chemistry B}
\bvolume{128}(\bissue{1}),
\bfpage{109}--\blpage{116}
(\byear{2024})
\doiurl{10.1021/acs.jpcb.3c06662}
{\href{https://arxiv.org/abs/https://doi.org/10.1021/acs.jpcb.3c06662}{{https://doi.org/10.1021/acs.jpcb.3c06662}}}.
\bcomment{PMID: 38154096}
\end{barticle}
\endbibitem

\bibitem[\protect\citeauthoryear{Anandakrishnan
  et~al.}{2015}]{Anandakrishnan_speedofconformations_2015}
\begin{barticle}
\bauthor{\bsnm{Anandakrishnan}, \binits{R.}},
\bauthor{\bsnm{Drozdetski}, \binits{A.}},
\bauthor{\bsnm{Walker}, \binits{R.}},
\bauthor{\bsnm{Onufriev}, \binits{A.}}:
\batitle{Speed of conformational change: Comparing explicit and implicit
  solvent molecular dynamics simulations}.
\bjtitle{Biophysical Journal}
\bvolume{108}(\bissue{5}),
\bfpage{1153}--\blpage{1164}
(\byear{2015})
\doiurl{10.1016/j.bpj.2014.12.047}
\end{barticle}
\endbibitem

\bibitem[\protect\citeauthoryear{Beauchamp et~al.}{2012}]{Beauchamp_ffnmr_2012}
\begin{barticle}
\bauthor{\bsnm{Beauchamp}, \binits{K.A.}},
\bauthor{\bsnm{Lin}, \binits{Y.-S.}},
\bauthor{\bsnm{Das}, \binits{R.}},
\bauthor{\bsnm{Pande}, \binits{V.S.}}:
\batitle{Are protein force fields getting better? a systematic benchmark on 524
  diverse nmr measurements}.
\bjtitle{Journal of Chemical Theory and Computation}
\bvolume{8}(\bissue{4}),
\bfpage{1409}--\blpage{1414}
(\byear{2012})
\doiurl{10.1021/ct2007814}
{\href{https://arxiv.org/abs/https://doi.org/10.1021/ct2007814}{{https://doi.org/10.1021/ct2007814}}}.
\bcomment{PMID: 22754404}
\end{barticle}
\endbibitem

\bibitem[\protect\citeauthoryear{Koes and Vries}{2017}]{Koes_ffnmrshifts_2017}
\begin{barticle}
\bauthor{\bsnm{Koes}, \binits{D.R.}},
\bauthor{\bsnm{Vries}, \binits{J.K.}}:
\batitle{Evaluating amber force fields using computed nmr chemical shifts}.
\bjtitle{Proteins: Structure, Function, and Bioinformatics}
\bvolume{85}(\bissue{10}),
\bfpage{1944}--\blpage{1956}
(\byear{2017})
\doiurl{10.1002/prot.25350}
{\href{https://arxiv.org/abs/https://onlinelibrary.wiley.com/doi/pdf/10.1002/prot.25350}{{https://onlinelibrary.wiley.com/doi/pdf/10.1002/prot.25350}}}
\end{barticle}
\endbibitem

\bibitem[\protect\citeauthoryear{Cavender
  et~al.}{2025}]{Cavender_structureff_2025}
\begin{botherref}
\oauthor{\bsnm{Cavender}, \binits{C.E.}},
\oauthor{\bsnm{Case}, \binits{D.A.}},
\oauthor{\bsnm{Chen}, \binits{J.C.-H.}},
\oauthor{\bsnm{Chong}, \binits{L.T.}},
\oauthor{\bsnm{Keedy}, \binits{D.A.}},
\oauthor{\bsnm{Lindorff-Larsen}, \binits{K.}},
\oauthor{\bsnm{Mobley}, \binits{D.L.}},
\oauthor{\bsnm{Ollila}, \binits{O.H.S.}},
\oauthor{\bsnm{Oostenbrink}, \binits{C.}},
\oauthor{\bsnm{Robustelli}, \binits{P.}},
\oauthor{\bsnm{Voelz}, \binits{V.A.}},
\oauthor{\bsnm{Wall}, \binits{M.E.}},
\oauthor{\bsnm{Wych}, \binits{D.C.}},
\oauthor{\bsnm{Gilson}, \binits{M.K.}}:
Structure-Based Experimental Datasets for Benchmarking Protein Simulation Force
  Fields
(2025).
\url{https://arxiv.org/abs/2303.11056}
\end{botherref}
\endbibitem

\bibitem[\protect\citeauthoryear{Bereau and
  Kremer}{2015}]{bereau2015automartini}
\begin{barticle}
\bauthor{\bsnm{Bereau}, \binits{T.}},
\bauthor{\bsnm{Kremer}, \binits{K.}}:
\batitle{Automated parametrization of the coarse-grained martini force field
  for small organic molecules}.
\bjtitle{J Chem Theory Comput}
\bvolume{11}(\bissue{6}),
\bfpage{2783}--\blpage{2791}
(\byear{2015})
\doiurl{10.1021/acs.jctc.5b00056}
\end{barticle}
\endbibitem

\bibitem[\protect\citeauthoryear{Kelidou et~al.}{2025}]{kelidou_automated_2025}
\begin{barticle}
\bauthor{\bsnm{Kelidou}, \binits{M.}},
\bauthor{\bsnm{Stroh}, \binits{K.S.}},
\bauthor{\bsnm{Risselada}, \binits{H.J.}}:
\batitle{Automated parametrization of small molecules within the {Martini} 3
  coarse-grained model guided by experimental log {P} values}.
\bjtitle{Scientific Reports}
\bvolume{15}(\bissue{1}),
\bfpage{37169}
(\byear{2025})
\doiurl{10.1038/s41598-025-24757-3}
\end{barticle}
\endbibitem

\bibitem[\protect\citeauthoryear{Smith et~al.}{2019}]{smith_approaching_2019}
\begin{barticle}
\bauthor{\bsnm{Smith}, \binits{J.S.}},
\bauthor{\bsnm{Nebgen}, \binits{B.T.}},
\bauthor{\bsnm{Zubatyuk}, \binits{R.}},
\bauthor{\bsnm{Lubbers}, \binits{N.}},
\bauthor{\bsnm{Devereux}, \binits{C.}},
\bauthor{\bsnm{Barros}, \binits{K.}},
\bauthor{\bsnm{Tretiak}, \binits{S.}},
\bauthor{\bsnm{Isayev}, \binits{O.}},
\bauthor{\bsnm{Roitberg}, \binits{A.E.}}:
\batitle{Approaching coupled cluster accuracy with a general-purpose neural
  network potential through transfer learning}.
\bjtitle{Nature Communications}
\bvolume{10}(\bissue{1}),
\bfpage{2903}
(\byear{2019})
\doiurl{10.1038/s41467-019-10827-4}
\end{barticle}
\endbibitem

\bibitem[\protect\citeauthoryear{Chen et~al.}{2023}]{Chen_demlpccsd_2023}
\begin{barticle}
\bauthor{\bsnm{Chen}, \binits{M.S.}},
\bauthor{\bsnm{Lee}, \binits{J.}},
\bauthor{\bsnm{Ye}, \binits{H.-Z.}},
\bauthor{\bsnm{Berkelbach}, \binits{T.C.}},
\bauthor{\bsnm{Reichman}, \binits{D.R.}},
\bauthor{\bsnm{Markland}, \binits{T.E.}}:
\batitle{Data-efficient machine learning potentials from transfer learning of
  periodic correlated electronic structure methods: Liquid water at afqmc,
  ccsd, and ccsd(t) accuracy}.
\bjtitle{Journal of Chemical Theory and Computation}
\bvolume{19}(\bissue{14}),
\bfpage{4510}--\blpage{4519}
(\byear{2023})
\doiurl{10.1021/acs.jctc.2c01203}
{\href{https://arxiv.org/abs/https://doi.org/10.1021/acs.jctc.2c01203}{{https://doi.org/10.1021/acs.jctc.2c01203}}}.
\bcomment{PMID: 36730728}
\end{barticle}
\endbibitem

\bibitem[\protect\citeauthoryear{Daru et~al.}{2022}]{Daru_ccsdmd_2022}
\begin{barticle}
\bauthor{\bsnm{Daru}, \binits{J.}},
\bauthor{\bsnm{Forbert}, \binits{H.}},
\bauthor{\bsnm{Behler}, \binits{J.}},
\bauthor{\bsnm{Marx}, \binits{D.}}:
\batitle{Coupled cluster molecular dynamics of condensed phase systems enabled
  by machine learning potentials: Liquid water benchmark}.
\bjtitle{Phys. Rev. Lett.}
\bvolume{129},
\bfpage{226001}
(\byear{2022})
\doiurl{10.1103/PhysRevLett.129.226001}
\end{barticle}
\endbibitem

\bibitem[\protect\citeauthoryear{Jumper et~al.}{2021}]{jumper_alphafold_2021}
\begin{barticle}
\bauthor{\bsnm{Jumper}, \binits{J.}},
\bauthor{\bsnm{Evans}, \binits{R.}},
\bauthor{\bsnm{Pritzel}, \binits{A.}},
\bauthor{\bsnm{Green}, \binits{T.}},
\bauthor{\bsnm{Figurnov}, \binits{M.}},
\bauthor{\bsnm{Ronneberger}, \binits{O.}},
\bauthor{\bsnm{Tunyasuvunakool}, \binits{K.}},
\bauthor{\bsnm{Bates}, \binits{R.}},
\bauthor{\bsnm{Žídek}, \binits{A.}},
\bauthor{\bsnm{Potapenko}, \binits{A.}},
\bauthor{\bsnm{Bridgland}, \binits{A.}},
\bauthor{\bsnm{Meyer}, \binits{C.}},
\bauthor{\bsnm{Kohl}, \binits{S.A.A.}},
\bauthor{\bsnm{Ballard}, \binits{A.J.}},
\bauthor{\bsnm{Cowie}, \binits{A.}},
\bauthor{\bsnm{Romera-Paredes}, \binits{B.}},
\bauthor{\bsnm{Nikolov}, \binits{S.}},
\bauthor{\bsnm{Jain}, \binits{R.}},
\bauthor{\bsnm{Adler}, \binits{J.}},
\bauthor{\bsnm{Back}, \binits{T.}},
\bauthor{\bsnm{Petersen}, \binits{S.}},
\bauthor{\bsnm{Reiman}, \binits{D.}},
\bauthor{\bsnm{Clancy}, \binits{E.}},
\bauthor{\bsnm{Zielinski}, \binits{M.}},
\bauthor{\bsnm{Steinegger}, \binits{M.}},
\bauthor{\bsnm{Pacholska}, \binits{M.}},
\bauthor{\bsnm{Berghammer}, \binits{T.}},
\bauthor{\bsnm{Bodenstein}, \binits{S.}},
\bauthor{\bsnm{Silver}, \binits{D.}},
\bauthor{\bsnm{Vinyals}, \binits{O.}},
\bauthor{\bsnm{Senior}, \binits{A.W.}},
\bauthor{\bsnm{Kavukcuoglu}, \binits{K.}},
\bauthor{\bsnm{Kohli}, \binits{P.}},
\bauthor{\bsnm{Hassabis}, \binits{D.}}:
\batitle{Highly accurate protein structure prediction with {AlphaFold}}.
\bjtitle{Nature}
\bvolume{596}(\bissue{7873}),
\bfpage{583}--\blpage{589}
(\byear{2021})
\doiurl{10.1038/s41586-021-03819-2}
\end{barticle}
\endbibitem

\bibitem[\protect\citeauthoryear{Watson et~al.}{2023}]{watson_novo_2023}
\begin{barticle}
\bauthor{\bsnm{Watson}, \binits{J.L.}},
\bauthor{\bsnm{Juergens}, \binits{D.}},
\bauthor{\bsnm{Bennett}, \binits{N.R.}},
\bauthor{\bsnm{Trippe}, \binits{B.L.}},
\bauthor{\bsnm{Yim}, \binits{J.}},
\bauthor{\bsnm{Eisenach}, \binits{H.E.}},
\bauthor{\bsnm{Ahern}, \binits{W.}},
\bauthor{\bsnm{Borst}, \binits{A.J.}},
\bauthor{\bsnm{Ragotte}, \binits{R.J.}},
\bauthor{\bsnm{Milles}, \binits{L.F.}},
\bauthor{\bsnm{Wicky}, \binits{B.I.M.}},
\bauthor{\bsnm{Hanikel}, \binits{N.}},
\bauthor{\bsnm{Pellock}, \binits{S.J.}},
\bauthor{\bsnm{Courbet}, \binits{A.}},
\bauthor{\bsnm{Sheffler}, \binits{W.}},
\bauthor{\bsnm{Wang}, \binits{J.}},
\bauthor{\bsnm{Venkatesh}, \binits{P.}},
\bauthor{\bsnm{Sappington}, \binits{I.}},
\bauthor{\bsnm{Torres}, \binits{S.V.}},
\bauthor{\bsnm{Lauko}, \binits{A.}},
\bauthor{\bsnm{De~Bortoli}, \binits{V.}},
\bauthor{\bsnm{Mathieu}, \binits{E.}},
\bauthor{\bsnm{Ovchinnikov}, \binits{S.}},
\bauthor{\bsnm{Barzilay}, \binits{R.}},
\bauthor{\bsnm{Jaakkola}, \binits{T.S.}},
\bauthor{\bsnm{DiMaio}, \binits{F.}},
\bauthor{\bsnm{Baek}, \binits{M.}},
\bauthor{\bsnm{Baker}, \binits{D.}}:
\batitle{De novo design of protein structure and function with {RFdiffusion}}.
\bjtitle{Nature}
\bvolume{620}(\bissue{7976}),
\bfpage{1089}--\blpage{1100}
(\byear{2023})
\doiurl{10.1038/s41586-023-06415-8}
\end{barticle}
\endbibitem

\bibitem[\protect\citeauthoryear{Kim et~al.}{2025}]{Kim_Multifidelity_2025}
\begin{barticle}
\bauthor{\bsnm{Kim}, \binits{J.}},
\bauthor{\bsnm{Kim}, \binits{J.}},
\bauthor{\bsnm{Kim}, \binits{J.}},
\bauthor{\bsnm{Lee}, \binits{J.}},
\bauthor{\bsnm{Park}, \binits{Y.}},
\bauthor{\bsnm{Kang}, \binits{Y.}},
\bauthor{\bsnm{Han}, \binits{S.}}:
\batitle{Data-efficient multifidelity training for high-fidelity machine
  learning interatomic potentials}.
\bjtitle{Journal of the American Chemical Society}
\bvolume{147}(\bissue{1}),
\bfpage{1042}--\blpage{1054}
(\byear{2025})
\doiurl{10.1021/jacs.4c14455}
{\href{https://arxiv.org/abs/https://doi.org/10.1021/jacs.4c14455}{{https://doi.org/10.1021/jacs.4c14455}}}.
\bcomment{PMID: 39688472}
\end{barticle}
\endbibitem

\bibitem[\protect\citeauthoryear{Zaverkin
  et~al.}{2023}]{Zaverkin_Transferlearning_2023}
\begin{barticle}
\bauthor{\bsnm{Zaverkin}, \binits{V.}},
\bauthor{\bsnm{Holzmüller}, \binits{D.}},
\bauthor{\bsnm{Bonfirraro}, \binits{L.}},
\bauthor{\bsnm{Kästner}, \binits{J.}}:
\batitle{Transfer learning for chemically accurate interatomic neural network
  potentials}.
\bjtitle{Phys. Chem. Chem. Phys.}
\bvolume{25},
\bfpage{5383}--\blpage{5396}
(\byear{2023})
\doiurl{10.1039/D2CP05793J}
\end{barticle}
\endbibitem

\bibitem[\protect\citeauthoryear{Zubatyuk
  et~al.}{2019}]{Zubatyuk_multitask_2019}
\begin{barticle}
\bauthor{\bsnm{Zubatyuk}, \binits{R.}},
\bauthor{\bsnm{Smith}, \binits{J.S.}},
\bauthor{\bsnm{Leszczynski}, \binits{J.}},
\bauthor{\bsnm{Isayev}, \binits{O.}}:
\batitle{Accurate and transferable multitask prediction of chemical properties
  with an atoms-in-molecules neural network}.
\bjtitle{Science Advances}
\bvolume{5}(\bissue{8}),
\bfpage{6490}
(\byear{2019})
\doiurl{10.1126/sciadv.aav6490}
{\href{https://arxiv.org/abs/https://www.science.org/doi/pdf/10.1126/sciadv.aav6490}{{https://www.science.org/doi/pdf/10.1126/sciadv.aav6490}}}
\end{barticle}
\endbibitem

\bibitem[\protect\citeauthoryear{Batatia
  et~al.}{2025}]{Batatia_foundationalmlp_2025}
\begin{barticle}
\bauthor{\bsnm{Batatia}, \binits{I.}},
\bauthor{\bsnm{Benner}, \binits{P.}},
\bauthor{\bsnm{Chiang}, \binits{Y.}},
\bauthor{\bsnm{Elena}, \binits{A.M.}},
\bauthor{\bsnm{Kovács}, \binits{D.P.}},
\bauthor{\bsnm{Riebesell}, \binits{J.}},
\bauthor{\bsnm{Advincula}, \binits{X.R.}},
\bauthor{\bsnm{Asta}, \binits{M.}},
\bauthor{\bsnm{Avaylon}, \binits{M.}},
\bauthor{\bsnm{Baldwin}, \binits{W.J.}},
\bauthor{\bsnm{Berger}, \binits{F.}},
\bauthor{\bsnm{Bernstein}, \binits{N.}},
\bauthor{\bsnm{Bhowmik}, \binits{A.}},
\bauthor{\bsnm{Bigi}, \binits{F.}},
\bauthor{\bsnm{Blau}, \binits{S.M.}},
\bauthor{\bsnm{Cărare}, \binits{V.}},
\bauthor{\bsnm{Ceriotti}, \binits{M.}},
\bauthor{\bsnm{Chong}, \binits{S.}},
\bauthor{\bsnm{Darby}, \binits{J.P.}},
\bauthor{\bsnm{De}, \binits{S.}},
\bauthor{\bsnm{Della~Pia}, \binits{F.}},
\bauthor{\bsnm{Deringer}, \binits{V.L.}},
\bauthor{\bsnm{Elijošius}, \binits{R.}},
\bauthor{\bsnm{El-Machachi}, \binits{Z.}},
\bauthor{\bsnm{Fako}, \binits{E.}},
\bauthor{\bsnm{Falcioni}, \binits{F.}},
\bauthor{\bsnm{Ferrari}, \binits{A.C.}},
\bauthor{\bsnm{Gardner}, \binits{J.L.A.}},
\bauthor{\bsnm{Gawkowski}, \binits{M.J.}},
\bauthor{\bsnm{Genreith-Schriever}, \binits{A.}},
\bauthor{\bsnm{George}, \binits{J.}},
\bauthor{\bsnm{Goodall}, \binits{R.E.A.}},
\bauthor{\bsnm{Grandel}, \binits{J.}},
\bauthor{\bsnm{Grey}, \binits{C.P.}},
\bauthor{\bsnm{Grigorev}, \binits{P.}},
\bauthor{\bsnm{Han}, \binits{S.}},
\bauthor{\bsnm{Handley}, \binits{W.}},
\bauthor{\bsnm{Heenen}, \binits{H.H.}},
\bauthor{\bsnm{Hermansson}, \binits{K.}},
\bauthor{\bsnm{Ho}, \binits{C.H.}},
\bauthor{\bsnm{Hofmann}, \binits{S.}},
\bauthor{\bsnm{Holm}, \binits{C.}},
\bauthor{\bsnm{Jaafar}, \binits{J.}},
\bauthor{\bsnm{Jakob}, \binits{K.S.}},
\bauthor{\bsnm{Jung}, \binits{H.}},
\bauthor{\bsnm{Kapil}, \binits{V.}},
\bauthor{\bsnm{Kaplan}, \binits{A.D.}},
\bauthor{\bsnm{Karimitari}, \binits{N.}},
\bauthor{\bsnm{Kermode}, \binits{J.R.}},
\bauthor{\bsnm{Kourtis}, \binits{P.}},
\bauthor{\bsnm{Kroupa}, \binits{N.}},
\bauthor{\bsnm{Kullgren}, \binits{J.}},
\bauthor{\bsnm{Kuner}, \binits{M.C.}},
\bauthor{\bsnm{Kuryla}, \binits{D.}},
\bauthor{\bsnm{Liepuoniute}, \binits{G.}},
\bauthor{\bsnm{Lin}, \binits{C.}},
\bauthor{\bsnm{Margraf}, \binits{J.T.}},
\bauthor{\bsnm{Magdău}, \binits{I.-B.}},
\bauthor{\bsnm{Michaelides}, \binits{A.}},
\bauthor{\bsnm{Moore}, \binits{J.H.}},
\bauthor{\bsnm{Naik}, \binits{A.A.}},
\bauthor{\bsnm{Niblett}, \binits{S.P.}},
\bauthor{\bsnm{Norwood}, \binits{S.W.}},
\bauthor{\bsnm{O’Neill}, \binits{N.}},
\bauthor{\bsnm{Ortner}, \binits{C.}},
\bauthor{\bsnm{Persson}, \binits{K.A.}},
\bauthor{\bsnm{Reuter}, \binits{K.}},
\bauthor{\bsnm{Rosen}, \binits{A.S.}},
\bauthor{\bsnm{Rosset}, \binits{L.A.M.}},
\bauthor{\bsnm{Schaaf}, \binits{L.L.}},
\bauthor{\bsnm{Schran}, \binits{C.}},
\bauthor{\bsnm{Shi}, \binits{B.X.}},
\bauthor{\bsnm{Sivonxay}, \binits{E.}},
\bauthor{\bsnm{Stenczel}, \binits{T.K.}},
\bauthor{\bsnm{Sutton}, \binits{C.}},
\bauthor{\bsnm{Svahn}, \binits{V.}},
\bauthor{\bsnm{Swinburne}, \binits{T.D.}},
\bauthor{\bsnm{Tilly}, \binits{J.}},
\bauthor{\bsnm{Oord}, \binits{C.}},
\bauthor{\bsnm{Vargas}, \binits{S.}},
\bauthor{\bsnm{Varga-Umbrich}, \binits{E.}},
\bauthor{\bsnm{Vegge}, \binits{T.}},
\bauthor{\bsnm{Vondrák}, \binits{M.}},
\bauthor{\bsnm{Wang}, \binits{Y.}},
\bauthor{\bsnm{Witt}, \binits{W.C.}},
\bauthor{\bsnm{Wolf}, \binits{T.}},
\bauthor{\bsnm{Zills}, \binits{F.}},
\bauthor{\bsnm{Csányi}, \binits{G.}}:
\batitle{A foundation model for atomistic materials chemistry}.
\bjtitle{The Journal of Chemical Physics}
\bvolume{163}(\bissue{18}),
\bfpage{184110}
(\byear{2025})
\doiurl{10.1063/5.0297006}
\end{barticle}
\endbibitem

\bibitem[\protect\citeauthoryear{Chen et~al.}{2026}]{chen_cgbg_2026}
\begin{botherref}
\oauthor{\bsnm{Chen}, \binits{W.}},
\oauthor{\bsnm{Zhao}, \binits{B.}},
\oauthor{\bsnm{Eckwert}, \binits{J.}},
\oauthor{\bsnm{Zavadlav}, \binits{J.}}:
Coarse-Grained Boltzmann Generators
(2026).
\url{https://arxiv.org/abs/2602.10637}
\end{botherref}
\endbibitem

\bibitem[\protect\citeauthoryear{Noé et~al.}{2019}]{noe_bg_2019}
\begin{barticle}
\bauthor{\bsnm{Noé}, \binits{F.}},
\bauthor{\bsnm{Olsson}, \binits{S.}},
\bauthor{\bsnm{Köhler}, \binits{J.}},
\bauthor{\bsnm{Wu}, \binits{H.}}:
\batitle{Boltzmann generators: Sampling equilibrium states of many-body systems
  with deep learning}.
\bjtitle{Science}
\bvolume{365}(\bissue{6457}),
\bfpage{1147}
(\byear{2019})
\doiurl{10.1126/science.aaw1147}
{\href{https://arxiv.org/abs/https://www.science.org/doi/pdf/10.1126/science.aaw1147}{{https://www.science.org/doi/pdf/10.1126/science.aaw1147}}}
\end{barticle}
\endbibitem

\bibitem[\protect\citeauthoryear{Fuchs et~al.}{2025}]{fuchs_2025_chemtrain}
\begin{barticle}
\bauthor{\bsnm{Fuchs}, \binits{P.}},
\bauthor{\bsnm{Thaler}, \binits{S.}},
\bauthor{\bsnm{Röcken}, \binits{S.}},
\bauthor{\bsnm{Zavadlav}, \binits{J.}}:
\batitle{chemtrain: Learning deep potential models via automatic
  differentiation and statistical physics}.
\bjtitle{Computer Physics Communications}
\bvolume{310},
\bfpage{109512}
(\byear{2025})
\doiurl{10.1016/j.cpc.2025.109512}
\end{barticle}
\endbibitem

\bibitem[\protect\citeauthoryear{Donnini
  et~al.}{2005}]{donnini_2005_incorporatingtheeffectofionicstrength}
\begin{barticle}
\bauthor{\bsnm{Donnini}, \binits{S.}},
\bauthor{\bsnm{Mark}, \binits{A.E.}},
\bauthor{\bsnm{Juffer}, \binits{A.H.}},
\bauthor{\bsnm{Villa}, \binits{A.}}:
\batitle{Incorporating the effect of ionic strength in free energy calculations
  using explicit ions}.
\bjtitle{Journal of Computational Chemistry}
\bvolume{26}(\bissue{2}),
\bfpage{115}--\blpage{122}
(\byear{2005})
\doiurl{10.1002/jcc.20156}
{\href{https://arxiv.org/abs/https://onlinelibrary.wiley.com/doi/pdf/10.1002/jcc.20156}{{https://onlinelibrary.wiley.com/doi/pdf/10.1002/jcc.20156}}}
\end{barticle}
\endbibitem

\bibitem[\protect\citeauthoryear{Halgren}{1996a}]{Halgren_mmff_1996}
\begin{barticle}
\bauthor{\bsnm{Halgren}, \binits{T.A.}}:
\batitle{Merck molecular force field. i. basis, form, scope, parameterization,
  and performance of mmff94}.
\bjtitle{Journal of Computational Chemistry}
\bvolume{17}(\bissue{5-6}),
\bfpage{490}--\blpage{519}
(\byear{1996})
\doiurl{10.1002/(SICI)1096-987X(199604)17:5/6<490::AID-JCC1>3.0.CO;2-P}
\end{barticle}
\endbibitem

\bibitem[\protect\citeauthoryear{Halgren}{1996b}]{Halgren_mmff2_1996}
\begin{barticle}
\bauthor{\bsnm{Halgren}, \binits{T.A.}}:
\batitle{Merck molecular force field. ii. mmff94 van der waals and
  electrostatic parameters for intermolecular interactions}.
\bjtitle{Journal of Computational Chemistry}
\bvolume{17}(\bissue{5-6}),
\bfpage{520}--\blpage{552}
(\byear{1996})
\doiurl{10.1002/(SICI)1096-987X(199604)17:5/6<520::AID-JCC2>3.0.CO;2-W}
\end{barticle}
\endbibitem

\bibitem[\protect\citeauthoryear{Halgren}{1996c}]{Halgren_mmff3_1996}
\begin{barticle}
\bauthor{\bsnm{Halgren}, \binits{T.A.}}:
\batitle{Merck molecular force field. iii. molecular geometries and vibrational
  frequencies for mmff94}.
\bjtitle{Journal of Computational Chemistry}
\bvolume{17}(\bissue{5-6}),
\bfpage{553}--\blpage{586}
(\byear{1996})
\doiurl{10.1002/(SICI)1096-987X(199604)17:5/6<553::AID-JCC3>3.0.CO;2-T}
\end{barticle}
\endbibitem

\bibitem[\protect\citeauthoryear{Halgren and
  Nachbar}{1996}]{Halgren_mmff4_1996}
\begin{barticle}
\bauthor{\bsnm{Halgren}, \binits{T.A.}},
\bauthor{\bsnm{Nachbar}, \binits{R.B.}}:
\batitle{Merck molecular force field. iv. conformational energies and
  geometries for mmff94}.
\bjtitle{Journal of Computational Chemistry}
\bvolume{17}(\bissue{5-6}),
\bfpage{587}--\blpage{615}
(\byear{1996})
\doiurl{10.1002/(SICI)1096-987X(199604)17:5/6<587::AID-JCC4>3.0.CO;2-Q}
\end{barticle}
\endbibitem

\bibitem[\protect\citeauthoryear{Halgren}{1996}]{Halgren_mmff5_1996}
\begin{barticle}
\bauthor{\bsnm{Halgren}, \binits{T.A.}}:
\batitle{Merck molecular force field. v. extension of mmff94 using experimental
  data, additional computational data, and empirical rules}.
\bjtitle{Journal of Computational Chemistry}
\bvolume{17}(\bissue{5-6}),
\bfpage{616}--\blpage{641}
(\byear{1996})
\doiurl{10.1002/(SICI)1096-987X(199604)17:5/6<616::AID-JCC5>3.0.CO;2-X}
\end{barticle}
\endbibitem

\bibitem[\protect\citeauthoryear{Bennett}{1976}]{Bennett_freeEnergy_1976}
\begin{barticle}
\bauthor{\bsnm{Bennett}, \binits{C.H.}}:
\batitle{Efficient estimation of free energy differences from monte carlo
  data}.
\bjtitle{Journal of Computational Physics}
\bvolume{22}(\bissue{2}),
\bfpage{245}--\blpage{268}
(\byear{1976})
\doiurl{10.1016/0021-9991(76)90078-4}
\end{barticle}
\endbibitem

\bibitem[\protect\citeauthoryear{Alvarez}{2021}]{alvarez_juwels_2021}
\begin{barticle}
\bauthor{\bsnm{Alvarez}, \binits{D.}}:
\batitle{{JUWELS} {Cluster} and {Booster}: {Exascale} {Pathfinder} with
  {Modular} {Supercomputing} {Architecture} at {Juelich} {Supercomputing}
  {Centre}}.
\bjtitle{Journal of large-scale research facilities JLSRF}
\bvolume{7},
\bfpage{183}
(\byear{2021})
\doiurl{10.17815/jlsrf-7-183} .
Accessed 2026-05-31
\end{barticle}
\endbibitem

\end{thebibliography}

\end{document}